
\input harvmac
\input tables
\input epsf

\overfullrule=0pt
 

\def\C{{\scriptscriptstyle C}}

\def\J{{\scriptscriptstyle J}}


\def\CF{{\cal F}}

\def\CH{{\cal H}}



\def\fourthirds{{4 \over 3}}
\def\half{{1 \over 2}}
\def\quarter{{1 \over 4}}
\def\sixth{{ 1\over 6}}
\def\third{{1 \over 3}}

\def\twothirds{{2 \over 3}}


\def\gtap{\raise.3ex\hbox{$>$\kern-.75em\lower1ex\hbox{$\sim$}}}

\def\ltap{\raise.3ex\hbox{$<$\kern-.75em\lower1ex\hbox{$\sim$}}}

\def\sp{\>\>}
\def\therefore{{\hbox{..}\kern-.43em \raise.5ex \hbox{.}}\>\>}

\def\betaL{\beta_L}
\def\phizero{\phi_0}
\def\Phizero{\Phi_0}
\def\bfH{{\bf H}}
\def\bfP{{\bf P}}
\def\Hsquid{{\bf H}_{\rm SQUID}}
\def\bfU{{\bf U}}
\def\bra#1{\langle #1 |}
\def\EC{E_\C}
\def\EJ{E_\J}
\def\ket#1{| #1 \rangle}
\def\dotproduct#1#2{\langle #1 | #2 \rangle}
\def\vev#1{\langle #1 \rangle}
\def\ccdot{\hbox{\kern-.1em$\cdot$\kern-.1em}}
\def\betavec{{\vec \beta} \, }
\def\mod{\, {\rm mod} \,}
\def\xvec{{\vec x} \, }
\def\xpvec{{\vec x}\, '}
\def\Xvec{\vec {\bf X}}
\def\pvec{{\vec p} \, }
\def\ppvec{{\vec p}\, '}
\def\Pvec{\vec {\bf P}}
\def\Jvec{\vec {\bf J}}
\def\Avec{\vec {\bf A}}
\def\nablavec{\vec {\nabla}}
\def\bfn{{\bf n}}
\def\bfphi{{\pmb \phi}}

\newdimen\pmboffset
\pmboffset 0.022em
\def\oldpmb#1{\setbox0=\hbox{#1}%
 \copy0\kern-\wd0
 \kern\pmboffset\raise 1.732\pmboffset\copy0\kern-\wd0
 \kern\pmboffset\box0}
\def\pmb#1{\mathchoice{\oldpmb{$\displaystyle#1$}}{\oldpmb{$\textstyle#1$}}
      {\oldpmb{$\scriptstyle#1$}}{\oldpmb{$\scriptscriptstyle#1$}}}

%


\nref\Feit{M.D. Feit, J.A. Fleck, Jr. and A. Steiger,
J. Comput. Phys. {\bf 47} (1982) 412.}
\nref\DeRaedt{H. DeRaedt, Comput. Phys. Rev. {\bf 7} (1987) 1.}
\nref\NumericalRecipes{W.H. Press. S.A. Teukolsky, W.T. Vetterling and
B.P. Flannery, {\it Numerical Recipes in C}, (Cambridge University Press,
1992) p.~851.}
\nref\BandraukI{A.D. Bandrauk and H. Shen, J. Chem. Phys. {\bf 99} (1993)
1185.} 
\nref\BandraukII{A.D. Bandrauk and H. Shen, J. Phys. A {\bf 27} (1994)
7147.}
\nref\Leforestier{C. Leforestier {\it et al.}, J. Comput. Phys. {\bf 94}
(1991) 59.}
\nref\Abramowitz{G. Blanch, ``Mathieu Functions'', in {\it Handbook of
Mathematical Functions}, ed. M. Abramowitz and I.A. Stegun (Dover
Publications, New York, 1970), p.~721.}
\nref\LeggettI{A.J. Leggett, Prog. Theor. Phys. Supp. {\bf 69} (1980) 80}
\nref\LeggettII{A.J. Leggett, Contemp. Phys. {\bf 25} (1984) 583.}
\nref\Friedman{J.R. Friedman, V. Patel, W. Chen, S.K. Tolpygo and
J.E. Lukens, Nature {\bf 406} (2000) 43.}
\nref\vanderWal{C.H. van der Wal, A.C.J. ter Haar, F.K. Wilhelm,
R.N. Schouten, C.J.P.M. Harmans, T.P. Orlando, S. Lloyd and J.E. Mooij,
Science {\bf 290} (2000) 773.}
\nref\Chiorescu{I. Chiorescu, Y. Nakamura, C.J.P.M. Harmons and J.E. Mooij,
Science {\bf 299} (2003) 1869.}
\nref\Josephson{B.D. Josephson, Phys. Lett. {\bf 1} (1962) 251.}
\nref\Feynman{R.P. Feynman, R.B. Leighton and M. Sands, {\it The Feynman
Lectures on Physics}, Vol III, Addison-Wesley (Reading) 1965, p. 21-8.}
\nref\Averin{D.V. Averin, Fortschritte der Physik, {\bf 48} (2000) 1055}
\nref\Makhlin{Y. Makhlin, G. Sch\"on and A. Shnirman, Rev. Mod. Phys. {\bf
73} (2001) 357.}
\nref\Crogan{M. Crogan, S. Khlebnikov and G. Sadiek,
Supercond. Sci. Technol, {\bf 15} (2002) 8.}
\nref\Corato{V. Corato, P. Silvestrini, L. Stodolsky and J. Wosiek,
Phys. Lett. A {\bf 309} (2003) 206.}
\nref\Berggren{K. Berggren, W. Oliver, J. Sage, P. Cho, in preparation.}


\nfig\cosinehundred{
Lowest ten energy eigenvalues for a quantum system coupled to a
$U(\beta)~=~2+2~\cos(2~\beta)$ potential with strength $\alpha=100$. The
potential's $\beta \to -\beta$ symmetry induces eigenstate pairing.  But
each pair's energy difference is not resolvable on this plot's scale.  The
first four excited pair energies are in the ratios 2.97, 4.89, 6.76 and
8.56 relative to the ground state energy.  Recall the corresponding
harmonic oscillator energy ratios are 3, 5, 7 and 9.}
\nfig\oneDharmonicosc{Low energy eigenvalues and eigenstates for 
a harmonic oscillator with coupling $\alpha=1$ to a quadratic potential.}
\nfig\twoDpotential{The two-dimensional periodic potential 
$U(\beta_x,\beta_y)=3 + \cos 2 \beta_y - 2 \cos \beta_x \cos \beta_y$.}
\nfig\twoDeigenstates{Lowest three energy eigenstates corresponding to the
periodic potential in \twoDpotential.}
\nfig\interference{Time evolution of ground plus first excited state
superposition.  The times in (a), (b) and (c) respectively correspond to
zero, one-quarter and one-half of the beat period.}
\nfig\packetposnspread{Mean position and width as functions of time  
for a one-dimensional Gaussian wavepacket (a) propagating in free space,
(b) hitting a wall and (c) bouncing around inside a box.}
\nfig\bouncingpacket{Probability density snapshots of a one-dimensional
Gaussian wavepacket bouncing around inside a box.}
\nfig\SQUIDcartoonI{An idealized SQUID pictured as a superconducting ring
carrying a supercurrent that is interrupted by an insulator section.}
\nfig\SQUIDcontour{Contour buried deep inside the
superconducting ring along which $\Jvec=0$.  The total magnetic flux
enclosed by the ring is proportional to the phase drop across the
junction.}
\nfig\SQUIDcircuit{
Equivalent circuit for a SQUID containing an inductor $L$ and Josephson
junction $J$.  The junction's capacitance is modeled by capacitor $C$.}
\nfig\SQUIDpotential{
SQUID potential corresponding to parameters $\beta_L=\phi_0=\pi$ and
coupling $\alpha=\EJ/\EC=10$. The colored horizontal lines denote the
system's lowest four energy eigenvalues.  The ground and first excited
states are so close together that their energies are not resolvable on this
plot's scale.}
\nfig\outputresponse{
The ``output'' response of a SQUID with $\beta_L=\phi_0=\pi$ to
the time varying coupling plotted in \alphavariation.}
\nfig\alphavariation{
The ``input'' time dependence of coupling $\alpha=\EJ/\EC$.}


\def\CITTitle#1#2#3{\nopagenumbers\abstractfont
\hsize=\hstitle\rightline{#1}
\vskip 0.0in\centerline{\titlefont #2} \centerline{\titlefont #3}
\abstractfont\vskip .2in\pageno=0}

\CITTitle{{\baselineskip=12pt plus 1pt minus 1pt
  \vbox{\hbox{WW-10438}\hbox{}\hbox{}}}}
{Low Energy Quantum System Simulation}{}
\centerline{Peter Cho and Karl Berggren
\footnote{}{This work is sponsored by the Air Force Office of Scientific 
Research under Air Force Contract F19628-00-C-002.  Opinions,
interpretations, conclusions, and recommendations are those of the authors
and are not necessarily endorsed by the United States Government.}}
\centerline{Lincoln Laboratory}
\centerline{Massachusetts Institute of Technology}
\centerline{Lexington, MA 02420}
 
\vskip .1in
\centerline{\bf Abstract}
\bigskip

A numerical method for solving Schr\"odinger's equation based upon a
Baker-Campbell-Hausdorff (BCH) expansion of the time evolution operator is
presented herein.  The technique manifestly preserves wavefunction norm,
and it can be applied to problems in any number of spatial dimensions.  We
also identify a particular dimensionless ratio of potential to kinetic
energies as a key coupling constant.  This coupling establishes
characteristic length and time scales for a large class of low energy
quantum states, and it guides the choice of step sizes in numerical work.
Using the BCH method in conjunction with an imaginary time rotation, we
compute low energy eigenstates for several quantum systems coupled to
non-trivial background potentials.  The approach is subsequently applied to
the study of 1D propagating wave packets and 2D bound state time
development.  Failures of classical expectations uncovered by simulations
of these simple systems help develop quantum intuition.

Finally, we investigate the response of a Superconducting Quantum
Interference Device (SQUID) to a time dependent potential.  We discuss how
to engineer the potential's energy and time scales so that the SQUID acts
as a quantum NOT gate.  The notional simulation we present for this gate
provides useful insight into the design of one candidate building block for
a quantum computer.

\Date{10/03}

\newsec{Introduction}

In this article, we study the low energy behavior of a quantum system that
interacts with an external classical background.  We specifically focus
upon numerically simulating the evolution of a system with just a few
degrees of freedom.  Its characteristic time scales are generally assumed
to be short compared with the background's, yet its typical velocities are
required to be low compared to the speed of light.  The dynamics of the
quantum system are then accurately described by the non-relativistic
Schr\"odinger equation
\foot{We work with units where $\hbar=1$.  We also use  bold-face symbols
to denote abstract quantum operators.}
\eqn\Schrodinger{
i {\partial \over \partial t} \ket{\psi} = {\bf H}(t) \ket{\psi}.}
Here the ``ket'' $\ket{\psi}$ denotes an abstract vector in a Hilbert space
that is in one-to-one correspondence with the physical state of the system.

The system's time evolution is governed by the Hamiltonian
\eqn\Hamiltonian{{\bf H}(t) = {\Pvec}^2 + V({\Xvec},t)}
which is a sum of kinetic and potential energy terms. We absorb an inverse
mass coefficient into ${\Pvec}^2$ so that the kinetic operator has unit
normalization.  The potential term summarizes the classical background's
effect on the quantum system.  The corresponding reaction of the background
to the system is ignored in Schr\"odinger's equation.  The time dependence
of $V$ consequently represents a fixed input rather than a dynamically
determined output.

If the potential is an arbitrary function of time, the Hamiltonian
evaluated at some time $t_1$ typically does not commute with the
Hamiltonian evaluated at another time $t_2$.  The formal solution to
Schr\"odinger's equation
\eqn\Schrodingersoln{\ket{\psi(t)} = \bfU(t) \ket{\psi(0)} }
then involves the unitary time evolution operator given by Dyson's formula
\eqna\Dyson
$$ \eqalignno{
\bfU(t) &= T \exp\Bigl[-i \int_0^t {\bf H}(t') dt' \Bigr] & \Dyson a \cr
&= 1 + \sum_{n=1}^\infty (-i)^n \int_0^t dt_1 \int_0^{t_1} dt_2 \cdots
\int_0^{t_{n-1}} dt_n \, \bfH(t_1) \bfH(t_2) \cdots \bfH(t_n). & \Dyson b\cr
} $$
The $T$ symbol in \Dyson{a}\ indicates that the exponential is time
ordered.  As the series expansion in \Dyson{b}\ explicitly demonstrates,
time ordering forces the temporally latest Hamiltonian factors to appear on
the left.  Given the complexity of \Dyson{}, it is little surprise that the
number of time dependent problems for which $\bfU(t)$ can be evaluated in
closed form is small.

If the classical background is independent of time, the evolution
operator reduces to
\eqn\timeindependentevolutionoperator{\bfU(t)= \exp(-i {\bf H} t).}
This exponentiated sum of non-commuting $\Pvec^2$ and $V(\Xvec)$ operators
can be decomposed into a product of exponentials via the
Baker-Campbell-Hausdorff (BCH) expansion:
\eqn\BCHdecomp{\bfU(t) = e^{O(t^4)} \times e^{-\third ( -{i t } )^3
\bigl[V(\Xvec),[V(\Xvec),\Pvec^2]\bigr] } e^{\sixth ( -{i t } )^3 
\bigl[[V(\Xvec),\Pvec^2], \Pvec^2 \bigr] } e^{-\half ( -{i t } )^2 
[V(\Xvec),\Pvec^2] } e^{-{i t} V(\Xvec)} 
e^{-{i t } \Pvec^2 }.}
Since the adjoint for each exponential factor equals its matrix inverse,
this decomposition manifestly preserves unitarity.

Although we are ultimately interested in simulating the response of quantum
systems to time varying backgrounds, we initially restrict our attention to
problems where the potential is not a function of time.  Expansion
\BCHdecomp\ then provides a practical means for evolving state vectors on a
computer.  For example, one readily finds
\eqna\evolutionformula
$$ \eqalignno{
\psi(\xvec,t) &= \dotproduct{\xvec}{\psi(t)} = e^{-i t V(\xvec)} \CF^{-1}
\Biggl( e^{-i t \pvec^2} \CF \Bigl( \psi(\xvec,0) \Bigr) \Biggr)
+ O(t^2) & \evolutionformula a \cr} $$
if just the leading exponential factors appearing on the right of
\BCHdecomp\ are retained.  Here $\CF$ and $\CF^{-1}$ denote Fourier and
inverse Fourier transform operations which are defined in Appendix~A.  If
the next-to-leading exponentiated commutator is kept in the BCH
decomposition of $\bfU(t)$, a more accurate evolution formula can be
derived using operator identities listed in Appendix~A:
$$ \eqalignno{
\psi(\xvec,t) &= e^{\half t^2 \nabla^2 V(\xpvec) -i t V(\xpvec)} \CF^{-1}
\Biggl( e^{-i t \pvec^2} \CF \Bigl( \psi(\xvec,0) \Bigr) \Biggr) 
+ O(t^3) & \evolutionformula b \cr} $$
with $\xpvec = \xvec + t^2 \nabla V(\xvec)$.  For our simulation work, we
include the next-to-next-to-leading exponentiated nested commutator factors
shown in eqn.~\BCHdecomp.  As the resulting generalization of
\evolutionformula{b}\ is complicated and un-illuminating, we relegate
it to Appendix~B.

Our numerical approach to solving Schr\"odinger's equation is a variant of
exponentiated split operator methods that have been explored by several
authors in the past \refs{\Feit,\DeRaedt}.  It can be implemented in any
number of spatial dimensions via efficient Fast Fourier Transform (FFT)
codes unlike other matrix inversion algorithms which become unwieldy beyond
one dimension \NumericalRecipes.  The price we pay for such generality is
having to Fourier transform back and forth between momentum and position
space.  As in all split operator approaches, such transforming is
computationally expensive.  So while others have studied more complicated
decompositions of $\bfU(t)$ that yield higher temporal order accuracy per
time step \refs{\BandraukI,\BandraukII}, we prefer to work with the simpler
evolution formulas in \evolutionformula{}\ and their subleading
generalizations which minimize the number of transforms that must be
performed.
\foot{According to the authors of ref.~\Leforestier\ who surveyed several 
different propagation schemes, it is unclear whether the gain in accuracy
achieved by higher order symmetric expansions of $\bfU(t)$ is offset by the
considerably greater numerical effort required to implement such schemes.}

The outline for our article is as follows.  In section~2, we identify a
particular energy ratio as a dimensionless coupling constant that
establishes characteristic length and time scales for several low energy
quantum systems.  For large values of this coupling, dimensional analysis
helps guide the choice of FFT bin sizes needed to implement our BCH
evolution algorithm.  In section~3, we discuss a simple technique based
upon an imaginary time rotation for projecting out low energy eigenstates
from trial wavefunctions.  The method is demonstrated to reproduce known
spectra for analytically soluble 1D models and to yield consistent
eigenfunctions for analytically intractable 2D systems.  In section~4, we
investigate Gaussian wavepacket propagation on three different backgrounds.
As we shall see, the quantum motion of a packet inside a box is, at first
glance, a surprise.  In section~5, we simulate the response of a
Superconducting Quantum Interference Device (SQUID) to a time varying
potential.  By judiciously modulating SQUID parameters, this device can be
forced to act as a quantum NOT gate.  Finally we summarize our findings in
section~6 and close with some thoughts on future simulation work for
quantum computer design.

\newsec{Naive dimensional analysis}

It is generally a good practice to scale out all dimensionful parameters
from Schr\"odinger's equation in order to gain qualitative insight into the
quantum system it describes.  Once dimensionful quantities have been
removed, Schr\"odinger's equation can only depend upon dimensionless ratios
whose values relative to unity encode non-trivial physics information.  In
particular, these ratios act as coupling constants which govern the
strength of the interaction between the quantum system and its classical
background.

In order to derive a dimensionless form of Schr\"odinger's equation, we first
restore all dimensionful parameters within Hamiltonian \Hamiltonian\ and
project it onto the position basis:
\eqn\posnbasisHamiltonian{
H(t) = {-\hbar^2 \over 2m} {\partial^2 \over \partial \xvec^2} +
V(\xvec,t).}
We next express the position vector $\xvec$ as the product of some
arbitrary unit of length $\ell$ and a dimensionless vector $\betavec$:
$$ \xvec \equiv \ell \betavec.$$
The energy $K_0 = \hbar^2 / 2 m \ell^2$ is subsequently scaled out from the
kinetic term in \posnbasisHamiltonian, and it is used to relate
dimensionful time $t$ to its dimensionless counterpart $\tau$:
$$ t = {\hbar \over K_0} \tau.$$
We also rewrite the potential term as another energy $V_0$ times a function
of $\betavec$ and $\tau$:
$$ V(\xvec,t) = V_0 \, U(\betavec,\tau). $$
The ratio 
\eqn\dimensionlessH{
\CH  \equiv {H \over K_0}  
= -{\partial^2 \over \partial \betavec^2} + \alpha \, U(\betavec,\tau)}
with $\alpha \equiv V_0/K_0$ then enters into the dimensionless Schr\"odinger
equation
$$ i {\partial \over \partial \tau} \psi(\betavec,\tau) = \CH(\tau)
\psi(\betavec,\tau). $$
We work with version \dimensionlessH\ of the Hamiltonian in all numerical
simulations.

If the classical background is complicated, it is generally characterized
by several length scales of different magnitudes.  The potential is then
implicitly a function of multiple length scale ratios:
$$ U = U\Bigl(\betavec,\tau; {L \over L'}, {L \over L''}, \cdots \Bigr). $$
Observable expectation values can depend in complicated ways upon $\alpha$
and these dimensionless ratios.  In this situation, naive dimensional
analysis cannot be used to establish physical properties of low energy
quantum systems such as their typical wavefunction spreads or oscillation
frequencies.

But for the important special case where the background is characterized by
a single length, dimensional considerations do fix the dependence of system
scales upon the $\alpha$ coupling.  For example, consider an oscillator
with Hamiltonian
\eqn\oscillatorH{\CH = -{\partial^2 \over \partial \betavec^2} 
+ \alpha \, (\betavec^2)^p. }
According to the virial theorem, the kinetic and potential energies of the
oscillator's low lying states are balanced on average:
\eqn\virial{\vev{K} = p \, \vev{V}.}
The states' characteristic length and time consequently scale with $\alpha$
as
\eqna\alphascaledependence
$$ \eqalignno{
L & \sim \bigl(\alpha p\bigr)^{-{1 \over 2p+2}}  \qquad\qquad
T \sim \bigl(\alpha p\bigr)^{-{1 \over p+1}},  
& \alphascaledependence a \cr} $$
while their momentum and energy are inversely related to $L$ and $T$:
$$ \eqalignno{
P & \sim \bigl(\alpha p\bigr)^{{1 \over 2p+2}}  \qquad\qquad
E \sim \bigl(\alpha p\bigr)^{{1 \over p+1}}.  
& \alphascaledependence b \cr} $$
Note in particular that $L \to 0$ as $\alpha \to \infty$.  As one would
intuitively expect, an oscillator's wavefunction becomes more tightly
concentrated about the potential's minimum as the potential's walls grow
more steep.

Consider next potentials such as $U=-\betavec^2+\betavec^4$ or
$U=\cos\betavec^2$ that possess non-vanishing quadratic terms at their
minima.  In the large $\alpha$ limit, self-consistent dimensional analysis
demonstrates that the scaling behavior in \alphascaledependence{a,b}\
continues to hold with power $p=1$.  As the wavefunction becomes
concentrated about the minimum point, it effectively senses only the
leading quadratic term in the potential's expansion about the minimum.  The
low energy spectra for a large class of systems therefore look like that
for a harmonic oscillator as $\alpha \to
\infty$.  As an illustrative example of this general behavior, we plot in
\cosinehundred\ the first ten eigenvalues for periodic energy eigenstates
of a one-dimensional cosine potential with $\alpha=100$.  Though the
eigenstates occur in nearly degenerate parity doublets, the familiar
harmonic oscillator level spacing is evident in the figure.

\midinsert
\centerline{
\epsfysize=7truecm \epsfbox[0 0 360 360]{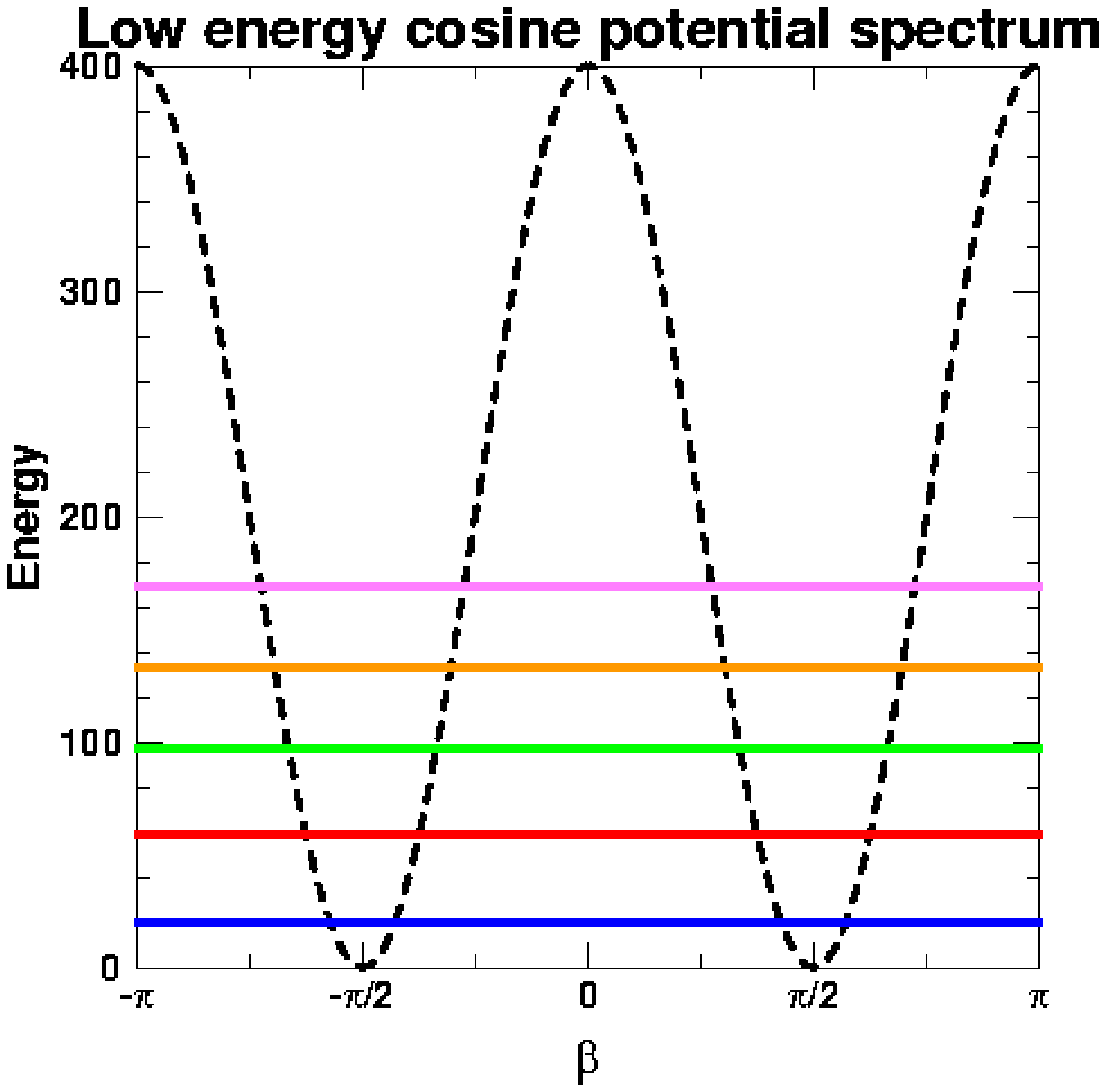}}
\bigskip
\hskip -0.6truecm\vbox{\hsize=16truecm  \noindent
Fig.~1: Lowest ten eigenvalues for periodic eigenstates of a quantum system
coupled to a $U(\beta)~=~2+2~\cos(2~\beta)$ potential with strength
$\alpha=100$.  The eigenstates occur in parity even and odd pairs, but each
pair's splitting is not resolvable on this plot's scale.  The first four
excited pair energies are in the ratios 2.97, 4.89, 6.76 and 8.56 relative
to the ground state energy.  Recall the corresponding harmonic oscillator
energy ratios are 3, 5, 7 and 9.}
\endinsert

These elementary dimensional analysis considerations help guide the
practical choice of FFT bin sizes needed to numerically implement the
Fourier transforms in eqns.~\evolutionformula{a,b}\ and their subleading
generalizations:
\eqn\binsizes{\delta \beta = {\alpha^{-1/4} \over \sqrt{N}} 
\qquad\qquad \delta \kappa = {\alpha^{1/4} \over \sqrt{N}}.}
In order to keep the position and momentum spaces on an equal footing, we
set both $\delta\beta$ and its dimensionless momentum analog $\delta
\kappa$ inversely proportional to the square root of the total number of
bins $N$ in each spatial dimension.  But we modulate the bins' magnitudes
by fractional powers of $\alpha$ to take into account generic systems'
length and momentum scales.  We likewise set the time step $\delta \tau$
used to evolve Schr\"odinger's equation proportional to $1/\sqrt{\alpha}$.
These scaling choices significantly improve simulation efficiency for
strongly coupled quantum systems.

\newsec{Energy eigenstate evaluation}

One of the most important properties of a quantum system is its low energy
spectrum.  It is natural to consider initially preparing a system in its
ground state or in a superposition of a few excited states.  We then want
to simulate how it evolves over time and reacts to changes in the classical
background.  In order to carry out this program, we first need to compute
the system's low energy eigenstates.

To begin, we center $U(\betavec)$ about the origin $\betavec=\vec{0}$.
Translating the potential is especially important for simulating
wavefunction evolution on periodic manifolds like $S^1$ (circle in 1D) or
$S^1 \times S^1$ (torus in 2D).  The periodicity of the FFT can be
exploited to numerically solve Schr\"odinger's equation on such manifolds
for states with periodic boundary conditions.  We therefore restrict the
domain for $\betavec$ to a single period in each spatial dimension for
which $U(\betavec)$ is periodic.

We next construct a trial wavefunction in position space for each low-lying
energy state.  In one spatial dimension, the parity of the ground state
corresponding to a symmetric potential $U(\beta)=U(-\beta)$ is always even.
Moreover, successive non-degenerate energy states labeled by principle
quantum number $n$ have alternating $(-1)^n$ parity assignments.  So for a
potential which is aperiodic in $\beta$, we take the $n^{th}$ eigenstate's
trial wavefunction to be proportional to $\beta^{(n \mod 2)}
\exp(-\beta^2)$.  If $U(\beta)$ is periodic on the other hand, we set
$\psi_{\rm trial}$ proportional to $\beta^{(n \mod 2)} U(\beta)^{-1}$.
These initial guesses are generally useful even when the potential does not
exhibit a perfect reflection symmetry.  Moreover, they correctly place the
bulk of low energy wavefunction content in valley regions of the potential.

Similar considerations motivate the choice of trial wavefunctions in higher
spatial dimensions.  For example in 2D, we set the trial wavefunction for
the energy eigenstate labeled by principle quantum numbers $m$ and $n$
proportional to $\beta_x^{(m \mod 2)} \beta_y^{(n \mod 2)}
U(\beta_x,\beta_y)^{-1}$ if the potential is periodic in both $\beta_x$ and
$\beta_y$.  On the other hand, we take $\psi_{\rm trial}$ to be a
two-dimensional Gaussian modulated by $\beta_x^{(m \mod 2)} \beta_y^{(n
\mod 2)}$ if the potential has infinite period.

Once a trial wavefunction for the ground state has been selected, it can be
written as an {\it a priori} unknown superposition of the true ground state
$\psi_0$ and all other energy eigenstates with which it shares the same
symmetry properties:
$$ \psi^{(0)}_{\rm trial}(\betavec,0) = C_0 \psi_0(\betavec,0) +
\sum_{j>0} C_j \psi_j(\betavec,0).$$
In order to distill $\psi_0$ from this infinite sum, we iteratively evolve
the trial state according to the imaginary time evolution operator $e^{-\CH
\tau}$ using subleading generalizations of eqn.~\evolutionformula{}\ with
$\tau \to -i \tau$:
$$ \psi^{(0)}_{\rm trial}(\betavec,\tau) = C_0 e^{-E_0 \tau}
\psi_0(\betavec,0) + \sum_{j>0} C_j e^{-E_j \tau} \psi_j(\betavec,0). $$
Following each timestep increment, the wavefunction is renormalized to
preserve amplitude content.  As the iteration proceeds, the trial
wavefunction's excited energy state components become exponentially
suppressed compared to the ground state term.  Ultimately, $\psi^{(0)}_{\rm
trial}$ relaxes to the true ground state
$$ \psi^{(0)}_{\rm trial} (\betavec,\tau)
\sp {\buildrel \tau \to \infty \over \longrightarrow} \sp
\psi_0(\betavec,0),  $$
and its Hamiltonian expectation value approaches the genuine lowest energy
eigenvalue:
$$ \bra{\psi^{(0)}_{\rm trial}(\tau)} \CH \ket{\psi^{(0)}_{\rm trial}(\tau) }
\sp {\buildrel \tau \to \infty \over \longrightarrow} \sp E_0 . $$

Once the ground state is known, we select a new trial wavefunction
$\psi^{(1)}_{\rm trial}$, compute its overlap with the lowest energy state
and form the difference
$$ \psi_{\rm reduced} = \psi^{(1)}_{\rm trial}-\dotproduct{\psi_0}{
\psi^{(1)}_{\rm trial}}\psi_0.$$  
We subsequently project out the first excited eigenstate from this reduced
wavefunction by repeating the imaginary time evolution procedure described
above for the ground state.  By following this projection and evolution
strategy, we can compute any energy eigenstate and eigenvalue so long as
build-up of numerical inaccuracies does not become overwhelmingly large.

As a first sanity check on this procedure, we illustrate its results in
\oneDharmonicosc\ for a 1D harmonic oscillator which couples with strength
$\alpha=1$ to a quadratic potential.  The numerically computed spectrum and
eigenfunctions reproduce well-known analytic formulas.  In order to recover
dimensionful energies from the dimensionless eigenvalues shown in the
figure, we need to multiply the latter by the energy $K_0 =
\half \hbar^2 / m \ell^2$ which we scaled out from Hamiltonian 
\dimensionlessH.  Recalling $V_0=\half m \omega^2 \ell^2$ for a simple
harmonic oscillator with natural frequency $\omega$ and $\alpha=V_0/K_0$,
we deduce $K_0 = \hbar \omega/2$.  So the $n^{th}$ state's dimensionless
$2n+1$ eigenvalue matches onto the dimensionful oscillator energy
$E_n=(2n+1)
\hbar \omega/2$.

\midinsert
\centerline{
\epsfysize=7truecm \epsfbox[0 0 684 358]{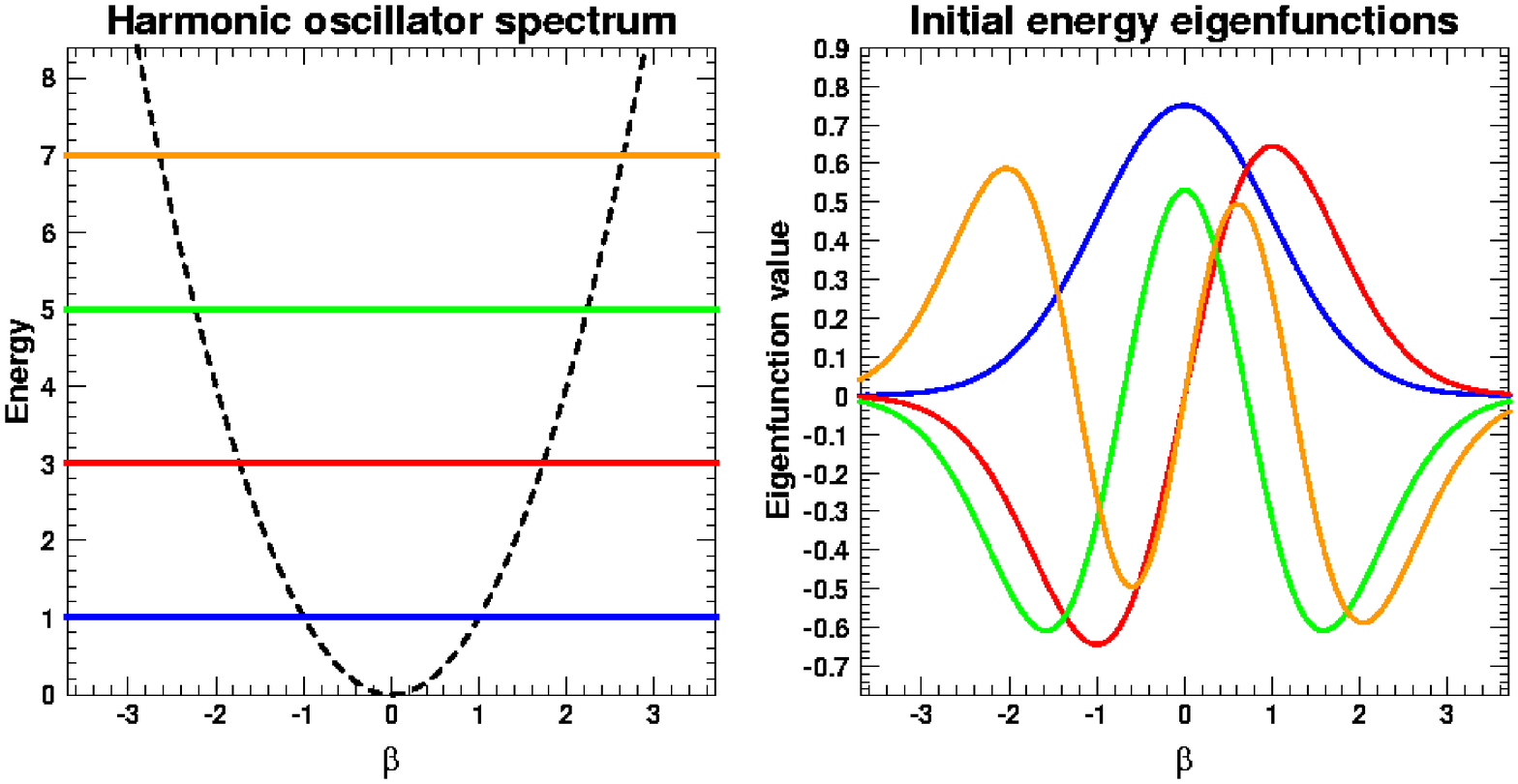}}
\smallskip
\hskip -0.6truecm\vbox{\hsize=16truecm  \noindent
Fig.~2: Low energy eigenvalues and eigenstates for a harmonic oscillator
with coupling $\alpha=1$ to a quadratic potential.}
\endinsert

As a more challenging test, we apply our eigenstate evaluation algorithm to
a system that interacts with potential $U(\beta)=2+2 \cos 2\beta$.  In this
case, the Schr\"odinger problem maps onto Mathieu's differential equation:
\eqn\mathieu{
\Bigl[ {\partial^2 \over \partial \beta^2} + (E-2 \alpha) - 2 \alpha 
\cos 2 \beta \Bigr] \psi(\beta) = 0.}
As we have have previously noted, the low energy spectrum for $\alpha
\gg 1$ looks like the harmonic oscillator's.  On the other hand, known power
series expressions for the eigenvalues of Mathieu's equation are convergent
when $\alpha \simeq O(1)$ \Abramowitz.  We therefore set $\alpha=1$ and
compare numerically derived eigenvalues with their power series
counterparts in Table~1. 

Some of the discrepancies between the analytic and numerical results shown
in the table are attributable to power series truncation.  In particular,
we estimate an $O(10^{-3})$ uncertainty in the zeroth and fourth analytic
eigenvalues based upon the magnitudes of the last power series terms listed
by Abramowitz and Stegun \Abramowitz.  Unfortunately, we have no simple way
to {\it a priori} quantify errors in our numerical eigenvalues.
Nevertheless, the agreement between the two sets of energies is overall
quite good for the first 11 periodic eigenstates.

\topinsert
\parasize=1in

\begintable
Energy \| Analytic \| Numerical \nr eigenstate \| eigenvalue \| eigenvalue
\crthick 0 \| 1.5457 \| 1.5434 \nr 1 \| 1.8897 \| 1.8897 \nr 2 \| 3.8591 \|
3.8587 \nr 3 \| 5.9170 \| 5.9170 \nr 4 \| 6.3704 \| 6.3826 \nr 5 \| 11.0477
\| 11.0477 \nr 6 \| 11.0784 \| 11.1114 \nr 7 \| 18.0330 \| 18.0329 \nr 8 \|
18.0338 \| 18.0958 \nr 9 \| 27.0208 \| 27.0209 \nr 10 \| 27.0209 \| 27.1204
\endtable
\bigskip
\hskip -0.6truecm\vbox{\hsize=16truecm  \noindent
Table.~1: Lowest 11 energy eigenvalues for a system with periodic
wavefunction boundary conditions that is coupled to potential
$U(\beta)~=~2+2~\cos(2~\beta)$ with strength $\alpha=1$.  The analytic
eigenvalues are based upon power series solutions to Mathieu's equation
\Abramowitz.}
\endinsert

As a final example, we consider a two-dimensional quantum system for which
no analytic spectrum solution is known.  It couples with unit strength to
the dimensionless potential
$$ U(\beta_x,\beta_y) = 3 + \cos 2\beta_y - 2 \cos\beta_x \cos\beta_y$$
plotted in \twoDpotential.  Note that $U$ remains invariant under $\beta_x
\leftrightarrow -\beta_x$ and $\beta_y \leftrightarrow -\beta_y$
reflections, but not under $\beta_x \leftrightarrow \beta_y$ exchange.  The
system's lowest three eigenvalues labeled by principle quantum numbers $m$
and $n$ equal $E_{00}=2.59$, $E_{01}=3.37$ and $E_{10}=3.79$.  Their
associated eigenfunctions $\psi_{00}$, $\psi_{01}$ and $\psi_{10}$ are
respectively displayed in figs.~4a, 4b and 4c.  As expected, the ground
state's amplitude is concentrated in valley regions of the 2D potential,
while its phase is everywhere uniform.  In contrast, the first and second
excited states exhibit node lines along the $\beta_y=0$ and $\beta_x=0$
axes, and their amplitudes on the two sides of these nodal separations are
$180^\circ$ out of phase.  Most of the excited states' wavefunction
densities avoid the potential's mountain terrain.

\midinsert
\centerline{
\epsfysize=7truecm \epsfbox[72 194  537 597]{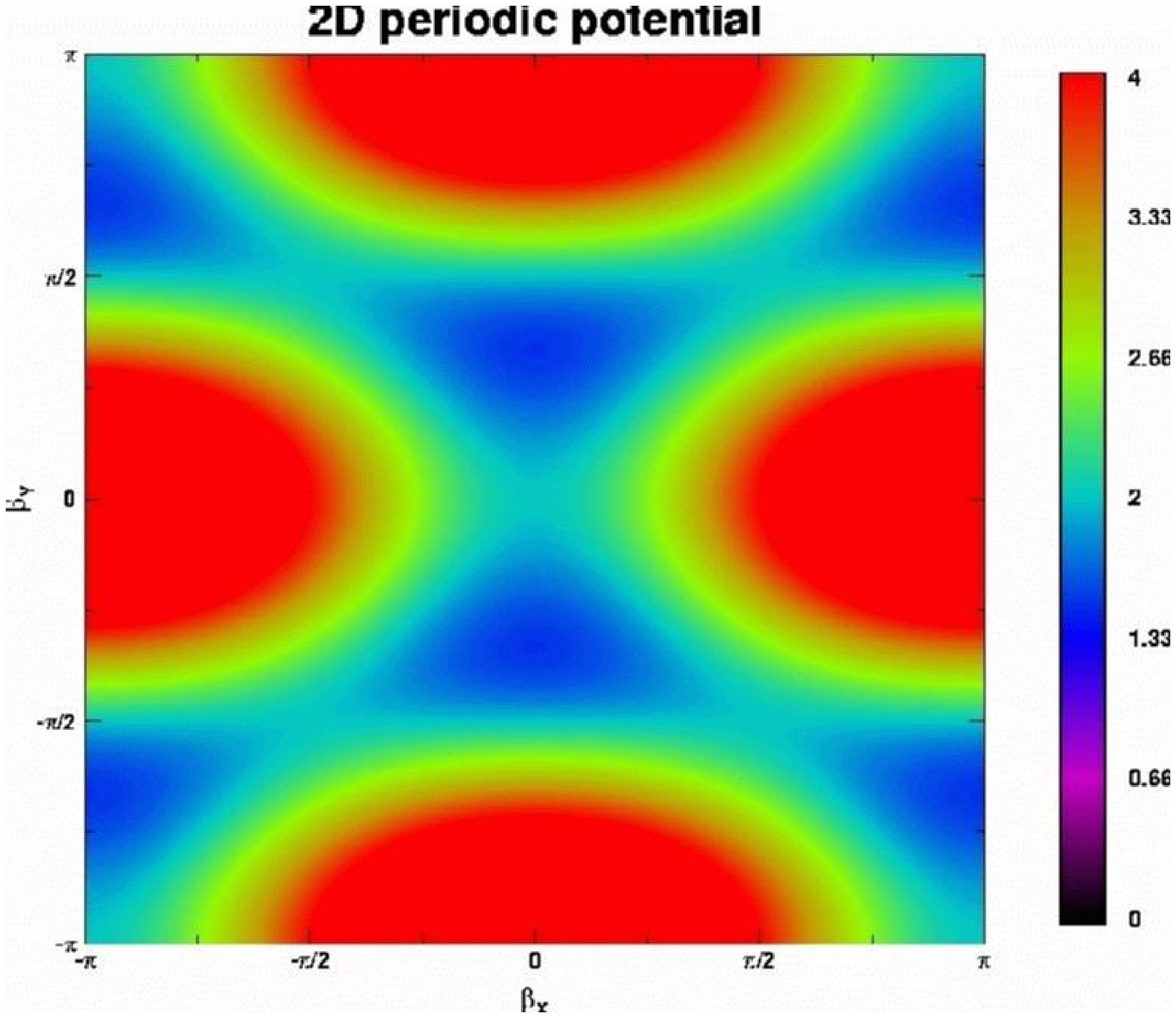}}
\bigskip
\hskip -0.6truecm\vbox{\hsize=16truecm  \noindent
Fig.~3: The two-dimensional periodic potential $U(\beta_x,\beta_y)=3 +
\cos 2 \beta_y - 2 \cos \beta_x \cos \beta_y$.}
\endinsert

The magnitudes and relative phases of these eigenfunctions must be
preserved as they evolve under the action of the unitary operator in
\timeindependentevolutionoperator.  
We have verified that these stationary wavefunction conditions are indeed
maintained up to small numerical errors.  Moreover, the states' energies
remain constant as time proceeds.  So although we cannot directly compare
the results in \twoDeigenstates\ with analytic expressions, our confidence
in their validity is high.

It is interesting to investigate time dependent interference between
combinations of the stationary states.  In \interference{a}, we display the
initial superposition
\eqn\superposition{\psi(\beta_x,\beta_y)={1 \over \sqrt 2} \Bigl( 
\psi_{00}(\beta_x,\beta_y) + \psi_{01}(\beta_x,\beta_y) \Bigr)}
of the ground and first excited states at $\tau=0$.  The dimensionless beat
period for this simple combination equals $T_{\rm beat} = 2
\pi/(E_{01}-E_{00})=8.1$.  In \interference{b} and \interference{c}, we plot
$\psi(\beta_x,\beta_y)$ at approximately the quarter and half way points
through the beat cycle.  Amplitude density tunneling between potential
valley regions for this superposition state is evident in the sequence of
wavefunction snapshots shown in the figure.

One could proceed to examine the time dependence of more complicated
superposition states.  But rather than continue the bound state discussion,
we turn at this point to investigate propagating states which exhibit
qualitatively different quantum phenomena.

\topinsert
\centerline{
\epsfysize=19truecm \epsfbox[73 100 537 689]{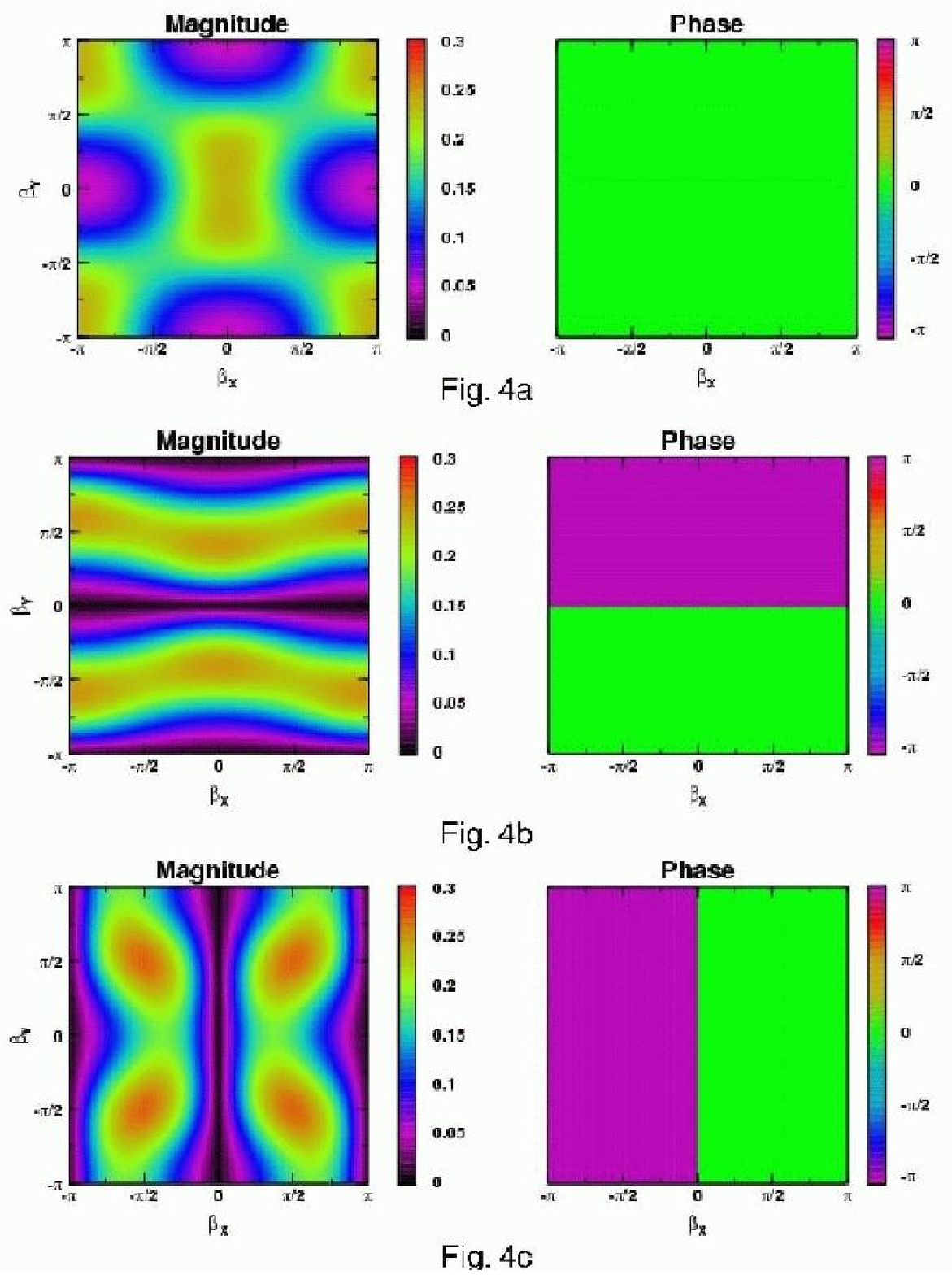}}
\bigskip\bigskip\bigskip
\hskip -0.6truecm\vbox{\hsize=16truecm  \noindent
Fig.~4: Lowest three energy eigenstates corresponding to the periodic
potential in \twoDpotential.}
\endinsert
\vfill\eject

\topinsert
\centerline{
\epsfysize=20truecm \epsfbox[73 100 537 689]{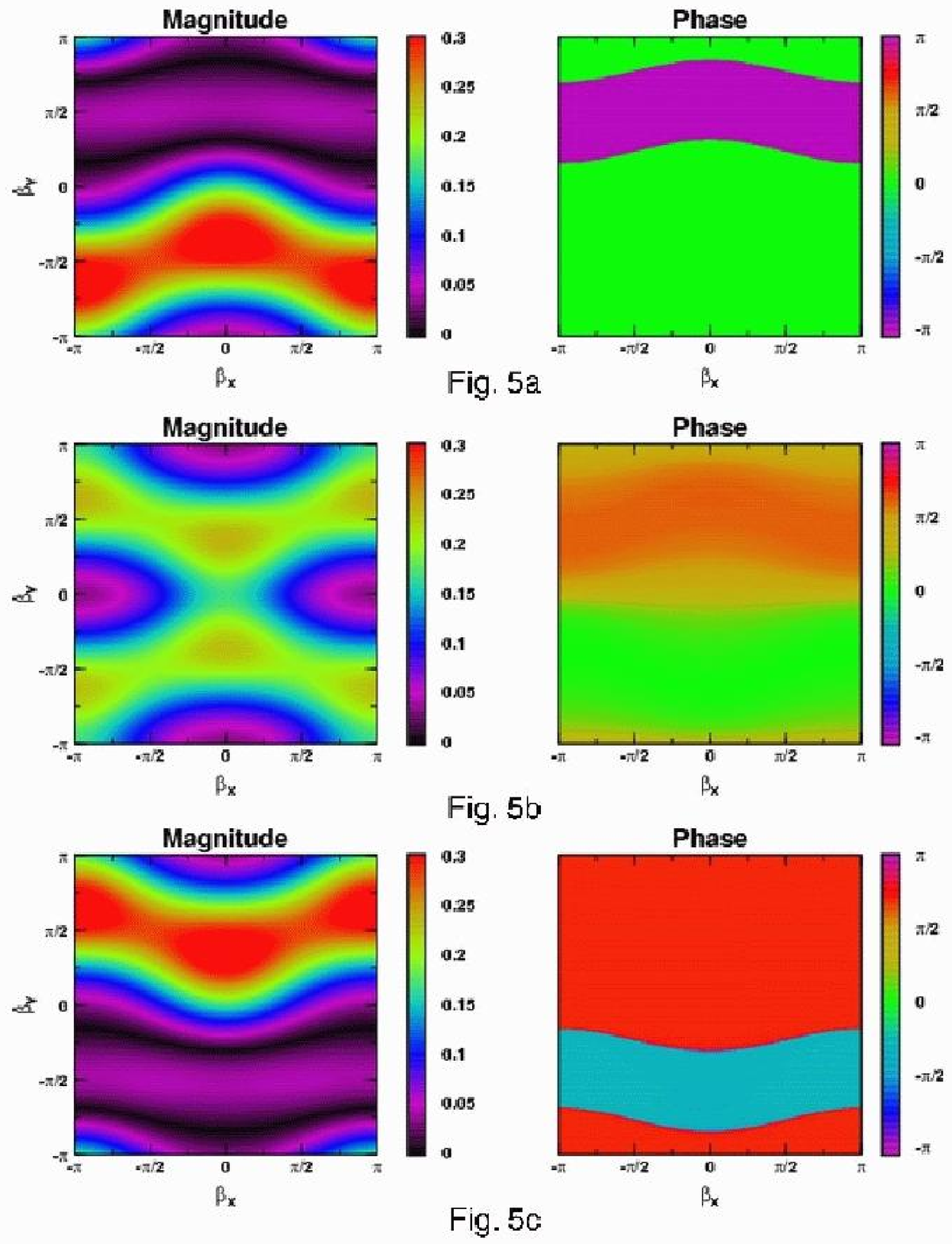}}
\bigskip\bigskip\bigskip
\hskip -0.6truecm\vbox{\hsize=16truecm  \noindent
Fig.~5: (a) Initial superposition of the ground and first excited states
from \twoDeigenstates.  (b) Superposition at $\tau \approx \quarter T_{\rm
beat}$.  (c) Superposition at $\tau \approx \half T_{\rm beat}$.}
\endinsert
\vfill\eject

\newsec{Gaussian wavepacket propagation}

We start our exploration of propagating states by considering the
well-known example of free Gaussian wavepacket evolution in one spatial
dimension.  The initial state is taken to have dimensionless mean position
$\beta_0$, dimensionless standard deviation $\sigma_0$ and dimensionless
momentum $k_0$:
\eqn\initialgaussian{
\psi(\beta,0) = \Bigl[ {1 \over 2\pi \sigma_0^2 } \Bigr]^{1/4}
\exp \Bigr[ - {\bigl( \beta - \beta_0)^2 \over 4 \sigma_0^2} \Bigr]
\exp \bigr[ i k_0 \beta \bigl].}
For the simple case of free propagation, Schr\"odinger's equation can be
solved analytically to determine the packet's time evolution:
\eqn\freegaussianpacket{
\psi(\beta,\tau) = \Bigl[ {1 \over 2\pi \sigma^2(\tau) } \Bigr]^{1/4}
\exp \Bigr[ - {\bigl( \beta - \beta_0 - 2 k_0 \tau \bigr)^2 \over
4 \sigma^2(\tau)} \Bigr] \times {(\rm phase \sp terms)}}
where 
\eqn\sigmaeff{\sigma(\tau) = \sqrt{\sigma_0^2 + \Bigl({\tau \over \sigma_0}
\Bigr)^2}.}
Though the wavefunction's phase terms are exactly calculable, we choose not
to explicitly list their ugly expressions here.

The mean motion of the packet
\eqn\meanmotion{\vev{\beta(\tau)} = \beta_0 + 2 k_0 \tau}
conforms with classical intuition.
\foot{Recall we have absorbed a factor of $1/2m$ into our
dimensionless Hamiltonian's kinetic term.  If all dimensionful factors are
restored, eqn.~\meanmotion\ reads $\vev{x(t)} = x_0 + p_0 t/m$.}
On the other hand, the packet's spreading over time is unexpected from
particle mechanics.  In order to demonstrate that the origin of this
diffusion is not purely a classical wave effect, we restore all
dimensionful parameters in \sigmaeff\ and expand the packet's width in
powers of $\hbar$:
\eqn\fullsigmaeff{s(t) = s_0 \sqrt{
1+{\hbar^2 t^2 \over 4 m^2 s_0^4}} = s_0 + O(\hbar^2).}
In the classical $\hbar \to 0$ limit, wavepacket diffusion does not occur.
So it represents a genuine quantum phenomenon.

These analytic results for free Gaussian packet evolution can be reproduced
by the numerical techniques described in the preceding sections.  For
example, we plot the mean position and standard deviation of a packet with
$\beta_0=0$, $\sigma_0=1/\sqrt{2}$ and $k_0=4$ as functions of time in
\packetposnspread{a}.  Numerical results shown in red are superposed for
comparison on analytic results shown in green.  Agreement between the two
is clearly quite good.

We consider next numerically propagating wave packets for problems where
closed form analytic formulas are difficult to obtain.  In
\packetposnspread{b}, the time dependent position and diffusion of 
a packet with the same dimensionless starting point, initial width and
momentum as that in \packetposnspread{a} are displayed.  However in this
second example, the packet encounters an infinite wall located at $\beta=8$
as represented by the dashed blue line in the mean position figure.  It
bounces off with no penetration into the energetically forbidden region.
During the time the packet interacts with the wall, the leading and
trailing portions of its wavefunction interfere with each other.  But after
the encounter is finished, the packet's mean motion once again looks
reasonably classical, and its diffusion is essentially unaffected by the
interaction with the wall.

The motion of a wavepacket inside a box is much more curious.  We again
take the initial state to be the same as that in
\packetposnspread{a}, and we graph its probability density  in 
the first snapshot of \bouncingpacket.  Subsequent snapshots in
\bouncingpacket\ depict the packet's evolution as it bounces between two
infinite walls positioned at $\beta=\pm 8$.  As time proceeds, the
diffusion of the packet becomes so pronounced that it continuously
interferes with itself everywhere inside the box.  Once the packet
completely fills the energetically allowed region, it no longer exhibits an
identifiable peak or momentum flow.  Consequently, its mean position decays
over time to zero (see \packetposnspread{c}).

One might wonder whether this last result violates Ehrenfest's theorem
which states that the laws of classical mechanics hold for quantum
expectation values:
\eqn\Ehrenfest{m {d^2  \over d t^2} \vev{x} = - {d \over d x}
\vev{V(x)}.}
But as $\vev{V(x)} \neq V(\vev{x})$, the wavepacket's mean position does
{\it not} obey Newton's law of motion.  So while a classical particle would
indefinitely ricochet off the walls of the box, the quantum wavepacket's
mean asymptotes instead to the center.  The motion of a quantum packet
trapped inside a box is thus highly non-classical.

\topinsert
\centerline{
\epsfysize=20truecm \epsfbox[72 101 538 688]{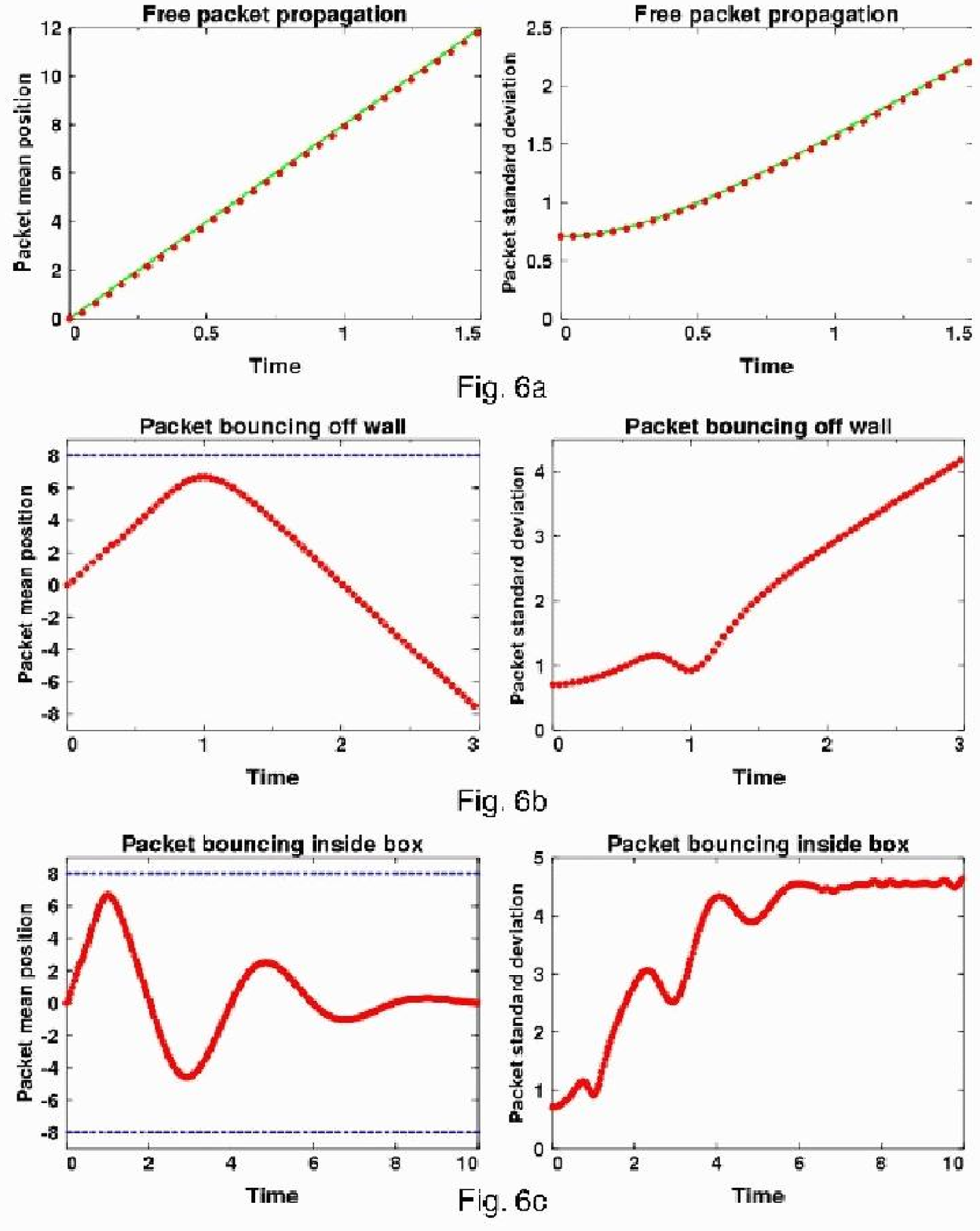}}
\bigskip\bigskip\bigskip
\hskip -0.6truecm\vbox{\hsize=16truecm  \noindent
Fig.~6: Mean position and standard deviation as functions of time for a 1D
Gaussian wavepacket (a) propagating in free space, (b) bouncing off a wall
and (c) bouncing inside a box.}
\endinsert
\vfill\eject

\topinsert
\centerline{
\epsfysize=19truecm \epsfbox[72 108 539 683]{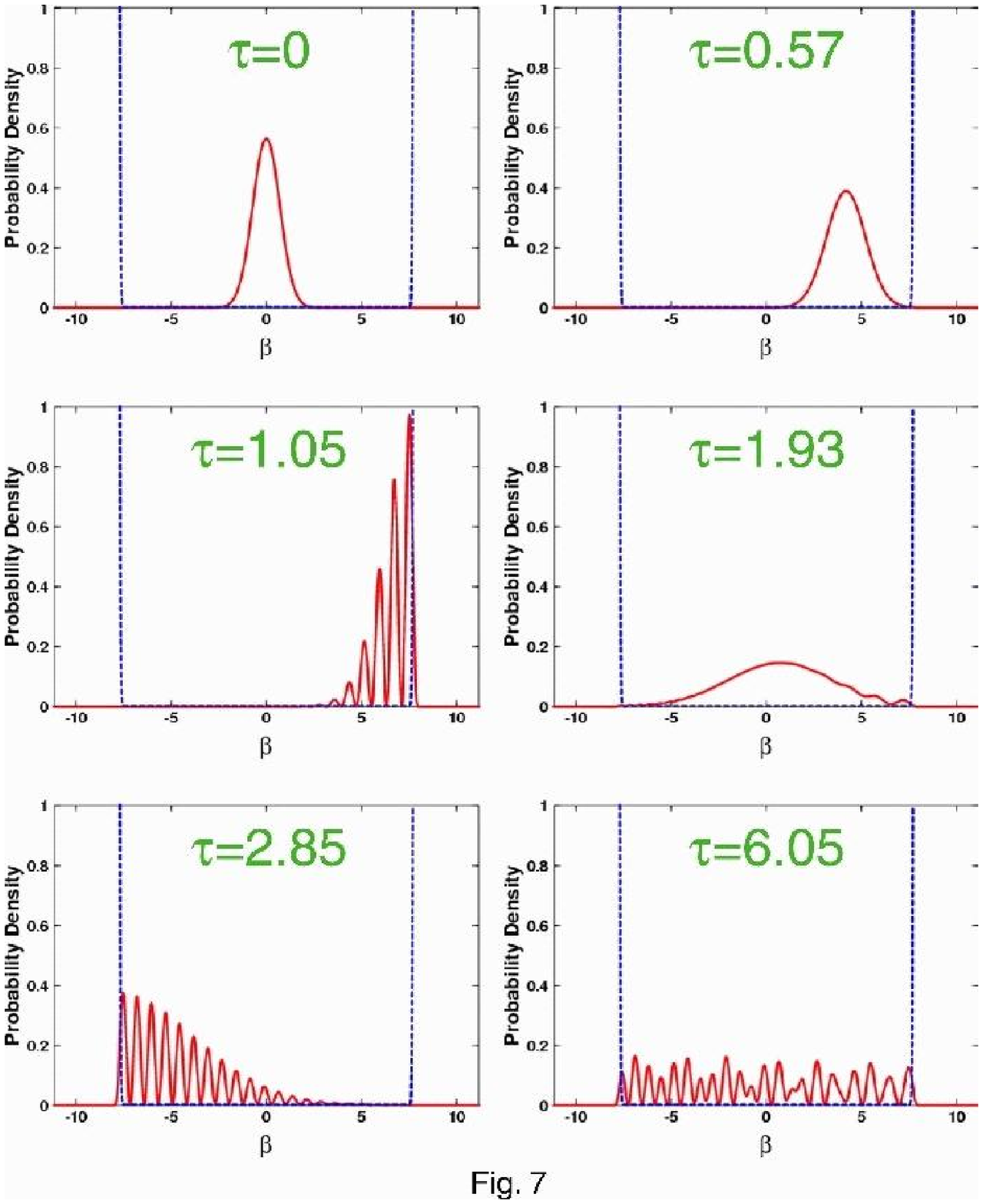}}
\bigskip\bigskip\bigskip
\hskip -0.6truecm\vbox{\hsize=16truecm  \noindent
Fig.~7: Probability density snapshots of a 1D Gaussian wavepacket bouncing
inside a box.}
\endinsert
\vfill\eject

\newsec{A notional quantum NOT gate}

We have so far investigated the interactions between simple quantum systems
with static classical backgrounds.  Now we consider modulating a potential
over time and measuring the system's response.  The spatial and temporal
dependence of potentials corresponding to various experimental setups may
be purposefully engineered to possess useful properties from an information
theory standpoint.  After a system is prepared in some initial state, its
wavefunction is intentionally manipulated via a time varying potential.
The system's final-state wavefunction is subsequently read out and measured
by the classical environment.  This general program is currently of great
interest for quantum computing applications.  In this section, we examine
the impact of a time dependent coupling upon one particular quantum system:
the SQUID.

Superconducting Quantum Interference Devices presently represent one of the
most promising candidate building blocks for quantum computers.  SQUID
circuits can theoretically support currents running clockwise and
counterclockwise simultaneously.  Recent experimental indications of such
superposition states are beginning to demonstrate the validity of quantum
mechanics on macroscopic scales \refs{\LeggettI,\LeggettII} as well as the
viability of SQUIDS as real-world qubits \refs{\Friedman{--}\Chiorescu}.
Moreover, SQUIDS are technologically attractive.  These passive devices are
small enough so that they can be mass produced on chips.  Yet they are
sufficiently large so that their quantum properties can be manipulated in a
controlled fashion.  SQUID design, fabrication and testing consequently
represent active areas of research.

As the quantum mechanics of SQUIDS may not be familiar to some readers, we
review the Hamiltonian $\Hsquid$ which governs their dynamics in
subsection~5.1.  The potential that enters into $\Hsquid$ looks like a
double-well within certain regions of its parameter space.  Wavefunction
localization in one of the two wells may naturally be interpreted as a
logical ``true'' or ``false'' signal.  If the couplings in $\Hsquid$ are
manipulated over time, coherent wavefunction movement from one well to the
other can be controlled.  So SQUID quantum mechanics has enough structure
for these devices to act as quantum NOT gates.

We subsequently explore SQUID dynamics in subsection~5.2 using our
simulation techniques.  It is important to note again that the naive
time dependent generalization of the evolution operator in
\timeindependentevolutionoperator
\eqn\naiveevolutionoperator{
\bfU(t)_{\rm naive} = \exp \Bigl[ -i \int_0^t \bfH(t') dt' \Bigr] }
differs from the true propagator in \Dyson{}\ by commutator terms which
enter at second order in the time step expansion:
\eqn\evolutionoperatordiff{
\bfU(t)_{\rm true} - \bfU(t)_{\rm naive} = \half \int_0^t dt_1 \int_0^{t_1}
dt_2 [ \bfH(t_1),\bfH(t_2) ] + O(t^3).}
These commutators render most numerical methods for solving Schr\"odinger's
equation less accurate for time dependent backgrounds, and ours provides no
exception to this general rule.  But since we are more interested in
rapidly gaining qualitative insight into SQUID experiment design than in
achieving high numerical precision, we continue to use
\evolutionformula{}\ and its subleading generalizations to propagate SQUID 
states interacting with time dependent potentials.  We do, however, employ
simple dynamic time step procedures to limit the accumulation of numerical
errors.

\subsec{The SQUID Hamiltonian}

To begin the derivation of $\Hsquid$, we sketch a cartoon of a
Superconducting Quantum Interference Device in \SQUIDcartoonI.  As the
figure illustrates, a SQUID is simply a superconducting ring interrupted by
a small insulator.  A supercurrent of Cooper pairs flows around the ring
and quantum mechanically tunnels through the junction.  The nonlinear
relationships between the insulator's current and voltage were worked out
by Josephson in 1962 \Josephson.  The insulator section acts as a nonlinear
circuit element, and it is referred to as a Josephson junction.

\midinsert
\centerline{
\epsfysize=6truecm \epsfbox[72 224 539 567 ]{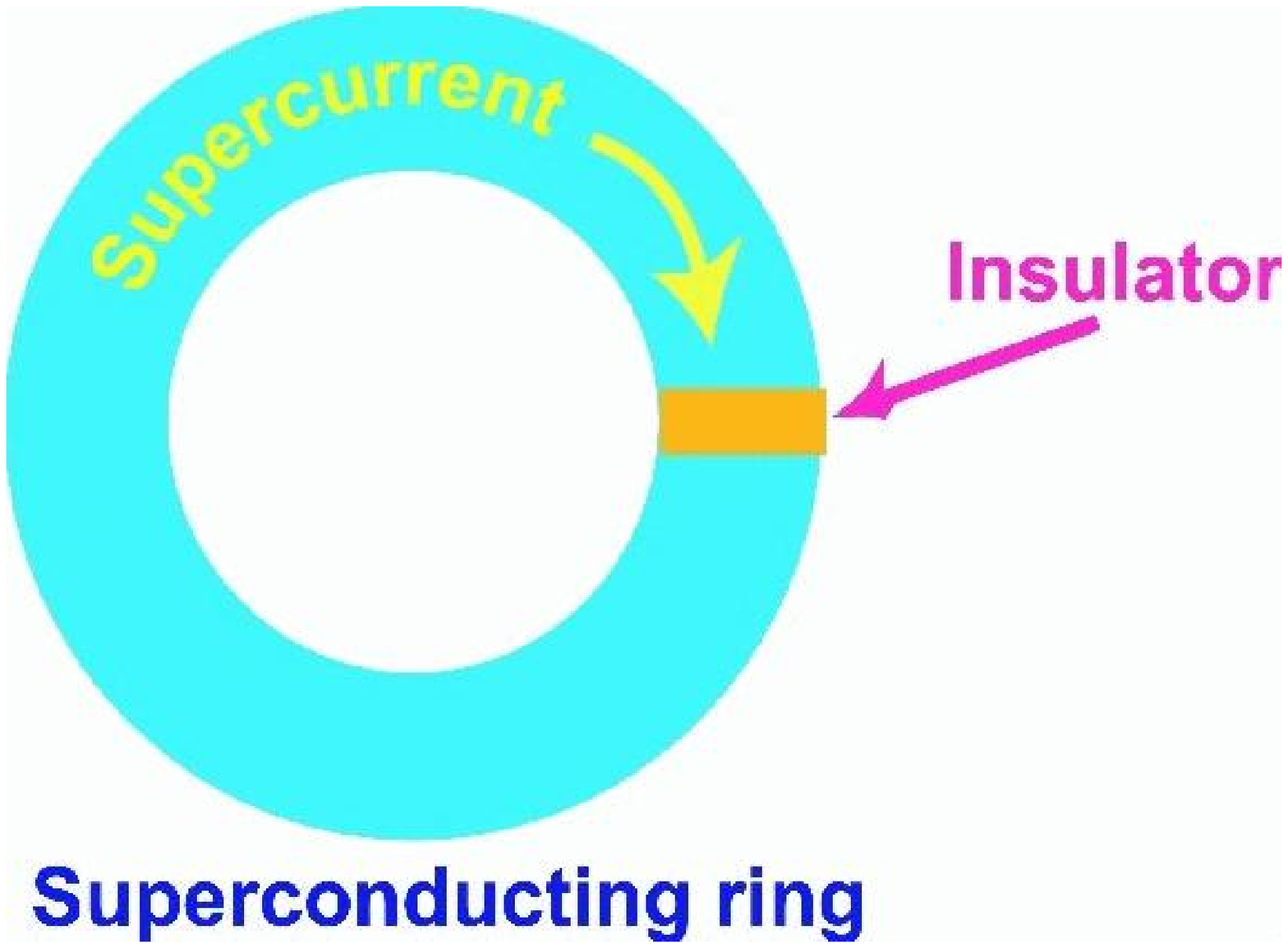}}
\bigskip
\hskip -0.6truecm\vbox{\hsize=16truecm  \noindent
Fig.~8: An idealized SQUID pictured as a superconducting ring
carrying a supercurrent that is interrupted by an insulator section.}
\endinsert

Following Feynman \Feynman, we let $\psi(\xvec)$ denote the wavefunction of
a low energy Cooper pair within the ring.  In the presence of a background
classical electromagnetic field, $\psi$ obeys a gauged Schr\"odinger
equation. Without loss of generality, it may be decomposed as $\psi =
\sqrt{\rho} e^{i \phi}$ where $\rho=\psi^* \psi$ is interpreted as 
Cooper pair probability density.  Since $\psi$ must be single-valued
everywhere along the ring, its phase $\phi$ must be a periodic function of
$\xvec.$

Probability density $\rho$ and its current counterpart $\Jvec$ satisfy a
gauged version of the familiar local conservation law $d\rho/dt = -
\nablavec \cdot \Jvec$.  This conservation requirement implies
$$ \Jvec = {1 \over 2m} \Bigl[ \psi^* \bigl( -i \hbar \nablavec \psi \bigr)
- \psi \bigl( -i \hbar \nablavec \psi^* \bigr) - 2 (2 e) \Avec \psi^* \psi
\Bigr] = {\hbar \over m} \Bigl[ \nablavec \phi - {2 e \over \hbar} \Avec
\Bigr] \rho $$
where $\Avec$ denotes the vector potential for the classical photon field.
As all currents within a superconductor are spatially localized near its
surface, $\Jvec$ vanishes deep inside its interior.  So
\eqn\phasereln{\nablavec \phi = {2 e \over \hbar} \Avec}
everywhere along a contour which lies buried in the SQUID ring (see
\SQUIDcontour).

\midinsert
\centerline{
\epsfysize=6truecm \epsfbox[72 157 539 633 ]{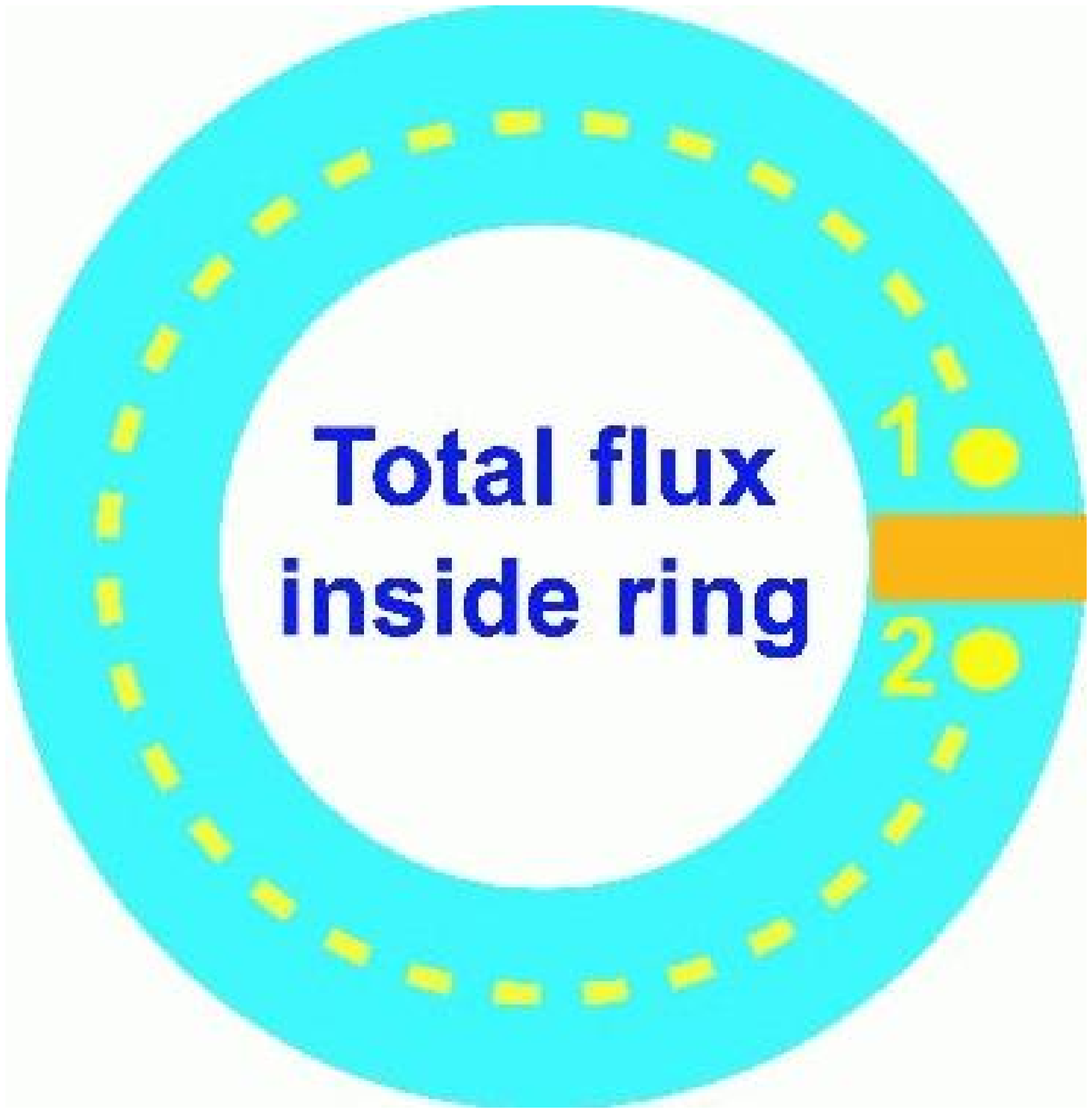}}
\bigskip
\hskip -0.6truecm\vbox{\hsize=16truecm  \noindent
Fig.~9: Contour deep inside the superconducting ring along which $\Jvec=0$.
The total magnetic flux enclosed by the ring is proportional to the phase
drop across the junction.}
\endinsert

The phase drop across the junction can be related to the total magnetic
flux that passes through the ring's hole by integrating \phasereln\ along
the contour between its endpoints:
\eqn\phasedrop{\eqalign{
\phi &\equiv \phi_2 - \phi_1 = \int_1^2 \nablavec \phi \cdot d\vec{s} \cr
&= {2e \over \hbar} \int_1^2 \Avec \cdot d\vec{s} 
   \simeq {2e \over \hbar} \oint \Avec \cdot d\vec{s} 
   = {2e \over \hbar} \int (\nablavec \times \Avec) \cdot d\vec{a} \cr
&= {2e \over \hbar} \Phi_{\rm total} 
= \Bigl({2\pi \over \Phizero}\Bigr) \Phi_{\rm total} .}}
The ratio $\Phizero \equiv h / 2 e$ represents a fundamental unit of flux.
Therefore, the junction phase drop counts the number of ``fluxons'' that
thread the ring.

The total magnetic flux appearing in \phasedrop\ is typically a sum of
supercurrent-generated internal flux plus some applied external flux:
$\Phi_{\rm total} = \Phi_{\rm int} + \Phi_{\rm ext}$.  The classical
expression for the energy associated with $\Phi_{\rm int}$ and the ring's
geometrical self-inductance $L$ motivates the first term in the SQUID's
quantum Hamiltonian:
$$ {\bf H}_{\rm inductance} = {{\bf \Phi}^2_{\rm int} \over 2L} 
=  {({\bf \Phi}_{\rm total}- {\Phi}_{\rm ext})^2 \over 2L} 
=  {(\Phizero/2 \pi)^2 \over 2L} ({\bfphi}^2 - \phi_0)^2. $$
In this formula, the phase offset is related to the external flux by
\phasedrop: $\phizero = (2\pi/\Phizero) \Phi_{\rm ext}$. 

A second contribution to $\Hsquid$ originates from the capacitance of the
Josephson junction.  The electric charge on the junction varies over time,
for not every Cooper pair tunnels through the insulator.  The classical
energy associated with this charge motivates the second Hamiltonian term
$$ {\bf H}_{\rm capacitance} = {{\bf Q}^2 \over 2C} = \EC \bfn^2. $$
The charge operator ${\bf Q} = 2 e \bfn$ effectively counts the number of
pairs on the junction.  When ${\bf H}_{\rm capacitance}$ is rewritten in
terms of the dimensionless number operator $\bfn$, its overall scale is set
by the Cooper pair capacitance energy $\EC=(2e)^2/2C$.

The Cooper pair number and phase operators are canonically conjugate to
each other, and they satisfy the commutation relation $[\bfn,\bfphi]=-i$.
In the $\phi$ representation, the number operator $\bfn \to -i
\partial/\partial \phi$ is seen to be the infinitesimal generator of $\phi$
translations.  Similarly, $\bfphi$ may be interpreted as the generator of
$\bfn$ translations.  These observations are important for understanding
the final contribution to the SQUID Hamiltonian
\eqn\Hjunction{{\bf H}_{\rm junction} = - {\EJ \over 2} 
\sum_{n=0}^\infty \Bigl\{ \ket{n+1} \bra{n} + \ket{n}\bra{n+1} \Bigr\} }
which describes the energy needed to transport a Cooper pair across the
Josephson junction \refs{\Averin,\Makhlin}.

The quantum expression in \Hjunction\ has no classical progenitor.  Its
prefactor is set by convention, while ${\bf S^+} \equiv \sum_{n=0}^\infty
\ket{n+1} \bra{n}$ and ${\bf S^-} = ({\bf S^+})^\dagger$ constitute
raising and lowering operators.  ${\bf S^\pm}$ may alternatively be
regarded as the finite translations $e^{\mp i \bfphi} \> {\buildrel n \>
{\rm rep} \over \longrightarrow} \> e^{\pm \partial/\partial n}$ which
alter pair number by $\pm 1$.  When rewritten in terms of these exponential
operators, the Josephson contribution reduces to its more familiar
sinusoidal form
$$ {\bf H}_{\rm junction} = -{\EJ \over 2} \Bigl[ e^{-i {\pmb \phi}} + e^{i
{\pmb \phi}} \Bigr] = - \EJ \cos {\pmb \phi}. $$

The sum of the inductance, capacitance and junction terms yields the SQUID
Hamiltonian
$$ \Hsquid = \EC \bfn^2 + { (\Phizero/2\pi)^2 \over 2L} (\bfphi-\phi_0)^2 -
\EJ \cos \bfphi . $$
The equations of motion following from this Hamiltonian
\eqna\squideom
$$ \eqalignno{
{d \bfn \over dt} &= {i \over \hbar} \bigl[ \Hsquid,\bfn \bigr] & \squideom
a \cr
{d \bfphi \over dt} &= {i \over \hbar} \bigl[ \Hsquid,\bfphi \bigr]
& \squideom b \cr } $$
are readily evaluated using the commutation relations listed in Appendix~A.
From \squideom{a}, we obtain a quantum version of Kirchoff's current
conservation law
\eqn\Kirchoff{{\bf I}_{\rm capacitance} + {\bf I}_{\rm inductance} + 
{\bf I}_{\rm junction} = 0}
with
\eqna\squidcurrents
$$ \eqalignno{
{\bf I}_{\rm capacitance} &= {d {\bf Q} \over dt} = 2 e {d \bfn \over dt} &
\squidcurrents a \cr
{\bf I}_{\rm inductance} &= {{\bf \Phi}_{\rm total} - \Phi_{\rm ext} \over L} 
= {(\Phizero/2 \pi) \over L} \bfphi & \squidcurrents b \cr
{\bf I}_{\rm junction} &= \Bigl( {2 \pi \over \Phizero} \Bigr) \EJ \sin
\bfphi \equiv I_c \sin \bfphi. & \squidcurrents c \cr} $$
Conservation relation \Kirchoff\ indicates that a SQUID can be modeled by
the equivalent circuit pictured in \SQUIDcircuit.  The cross appearing in
the figure represents the Josephson junction and its supercurrent flow
\squidcurrents{c}\ which depends upon the phase drop across the junction.

\midinsert
\centerline{
\epsfysize=5truecm \epsfbox[70 280 540 510 ]{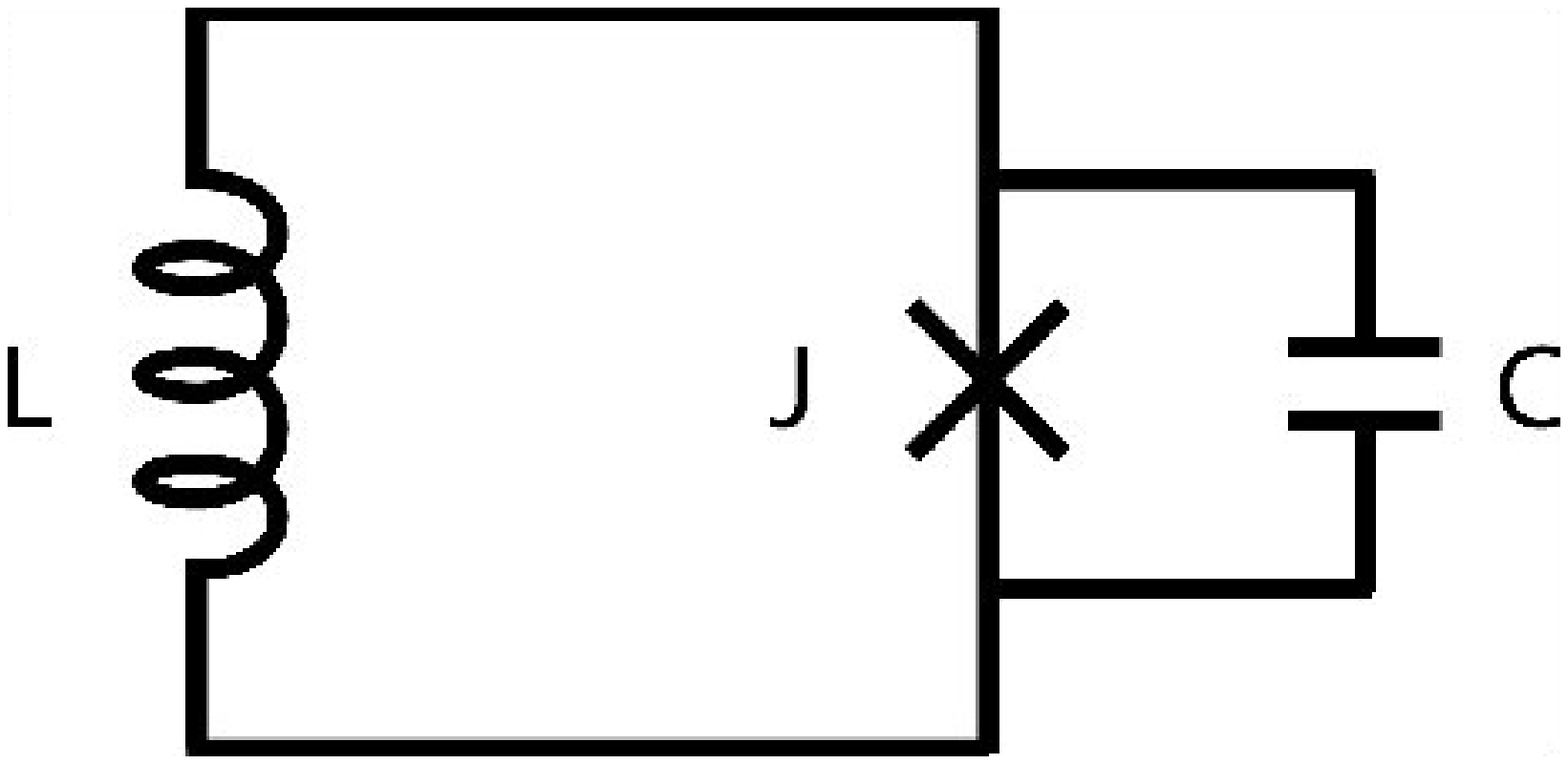}}
\bigskip
\hskip -0.6truecm\vbox{\hsize=16truecm  \noindent
Fig.~10: Equivalent circuit for a SQUID containing an inductor $L$ and
Josephson junction $J$.  The junction's capacitance is modeled by capacitor
$C$.}
\endinsert

So long as $I_{\rm junction}$ remains less than the critical value $I_c
\equiv (2\pi/\Phizero) \EJ$, the supercurrent tunnels through the insulator
without dissipation.  The voltage counterpart to $I_{\rm junction}$ also
depends upon the phase drop $\phi$, and it is obtained from the second
equation of motion in \squideom{b}:
\eqn\junctionvoltage{{\bf V}_{\rm junction} = \Bigl({\Phizero \over 2 \pi}
\Bigr) {d \bfphi \over dt}. }
The Josephson junction's I-V characteristics are governed by
eqns.~\squidcurrents{c}\ and \junctionvoltage.

We make two final comments about $\Hsquid$.  Firstly, it is customary to
introduce the dimensionless parameter
$$ \betaL \equiv {L I_c \over \bigl( \Phizero/2 \pi \bigr)} = L \Bigl( {2
\pi \over \Phizero} \Bigr)^2 \EJ $$
which counts the number of ``fluxons'' that thread the SQUID ring when it
carries critical current $I_c$.  When parameter $L$ is eliminated in favor
of $\betaL$, the Hamiltonian in the $\phi$ representation reduces to
\eqn\HSQUIDtwo{H_{\rm SQUID} = -\EC {\partial^2 \over \partial \phi^2} + 
\EJ \Bigl[ {(\phi-\phi_0)^2 \over 2 \betaL} - \cos\phi \Bigr]. }
Secondly, we scale out capacitance energy $\EC$ from $H_{\rm SQUID}$ for
simulation purposes.  The dimensionless SQUID Hamiltonian then takes the
same form as the template in \dimensionlessH
\eqn\dimenionslessHsquid{
\CH = - {\partial^2 \over \partial \phi^2} + \alpha U(\phi)}
with coupling constant $\alpha=\EJ/\EC$ and potential
\eqn\VSQUID{U = {(\phi-\phizero)^2 \over 2 \betaL} - \cos \phi.}
%

\subsec{SQUID response to a time dependent background
\foot{After work on this article was completed, we learned that a similar
analysis of SQUID response to a time dependent pulse was recently reported
in ref.~\Crogan.  Where overlap exists, there is generally good agreement
between the findings of ref.~\Crogan\ and the results independently derived
in this subsection.}}
%

The SQUID Hamiltonian is a function of the two free parameters $\betaL$ and
$\phizero$ in addition to the overall coupling $\alpha=\EJ/\EC$.  If naive
dimensional analysis intuition is to hold, $\betaL$ and $\phizero$ must
both be of order unity.  Their precise values control the potential's shape
and thereby strongly influence one's ability to extract useful information
from low lying quantum states.  We choose to fix $\betaL=\phizero=\pi$ so
that the SQUID potential looks like a double-well centered about
$\phi=\phizero$.  For simulation as well as visualization purposes, we also
initially set $\alpha=10$.  $U(\phi)$ is plotted for these parameter
choices in \SQUIDpotential.

\midinsert
\centerline{
\epsfysize=7truecm \epsfbox[0 0 580 450 ]{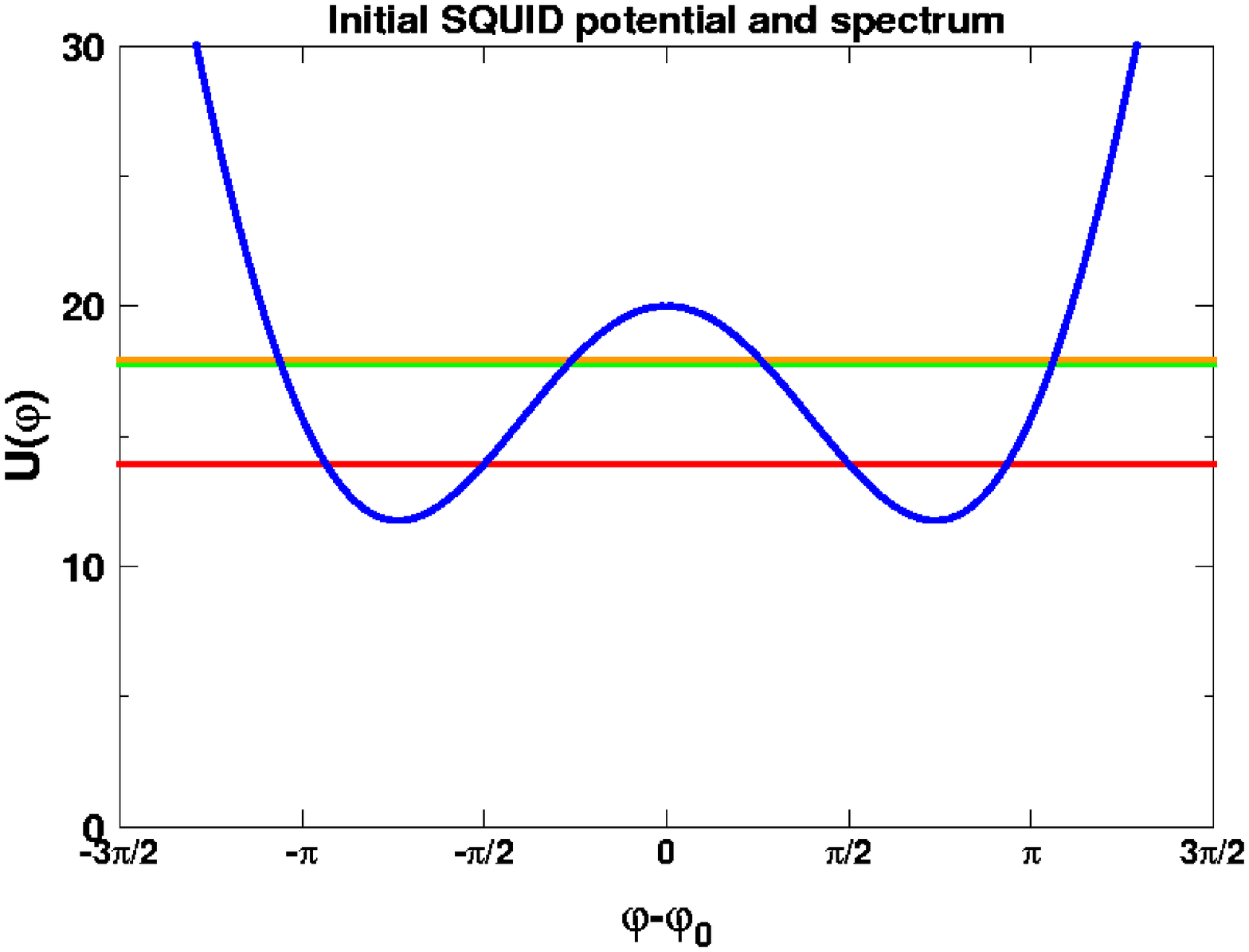}}
\bigskip
\hskip -0.6truecm\vbox{\hsize=16truecm  \noindent
Fig.~11: Dimensionless SQUID potential corresponding to parameters
$\betaL=\phizero=\pi$ and coupling $\alpha=\EJ/\EC=10$. The colored
horizontal lines denote the system's lowest four energy eigenvalues.  
The splitting between the ground and first excited state energies is not
resolvable on this plot's energy scale.}
\endinsert

The Hamiltonian remains invariant under a discrete $\phi-\pi \to
-(\phi-\pi)$ symmetry, and the system's energy eigenstates are even and odd
with respect to this parity operation.  As the horizontal energy lines in
\SQUIDpotential\ demonstrate, the low lying parity partners are nearly
degenerate:
\eqn\initSQUIDspectrum{\eqalign{
E_0^{(+)} = 13.8916   \qquad    E_0^{(-)} = 13.8960 \cr
E_1^{(+)} = 17.7426   \qquad    E_1^{(-)} = 17.9174. \cr}}
These energy eigenvalues establish relevant time scales for the design of a
quantum NOT gate.  For example, suppose the SQUID is initially prepared in
the state $\psi = (\psi_0^{(+)} + \psi_0^{(-)})/\sqrt{2}$ which is
localized in the positive well and corresponds to a counterclockwise
current.  In half a beat period $0.5 \, T_{\rm beat}= \pi / (E_0^{(-)} -
E_0^{(+)}) = 714$, the system naturally tunnels into the negative well, and
the current flows in a clockwise direction.  Intentional switching of the
SQUID's wavefunction must obviously be performed and measured on a much
shorter time scale.
\foot{In this notional NOT gate example, we make no attempt to model
decoherence effects which are present in any real-world quantum experiment.
But it is important to realize that the decoherence time could well set a
much more stringent upper bound on when a measurement must be performed
than does $0.5 \, T_{\rm beat}$.}
\topinsert
\centerline{
\epsfysize=19.75truecm \epsfbox[76 72 535 719 ]{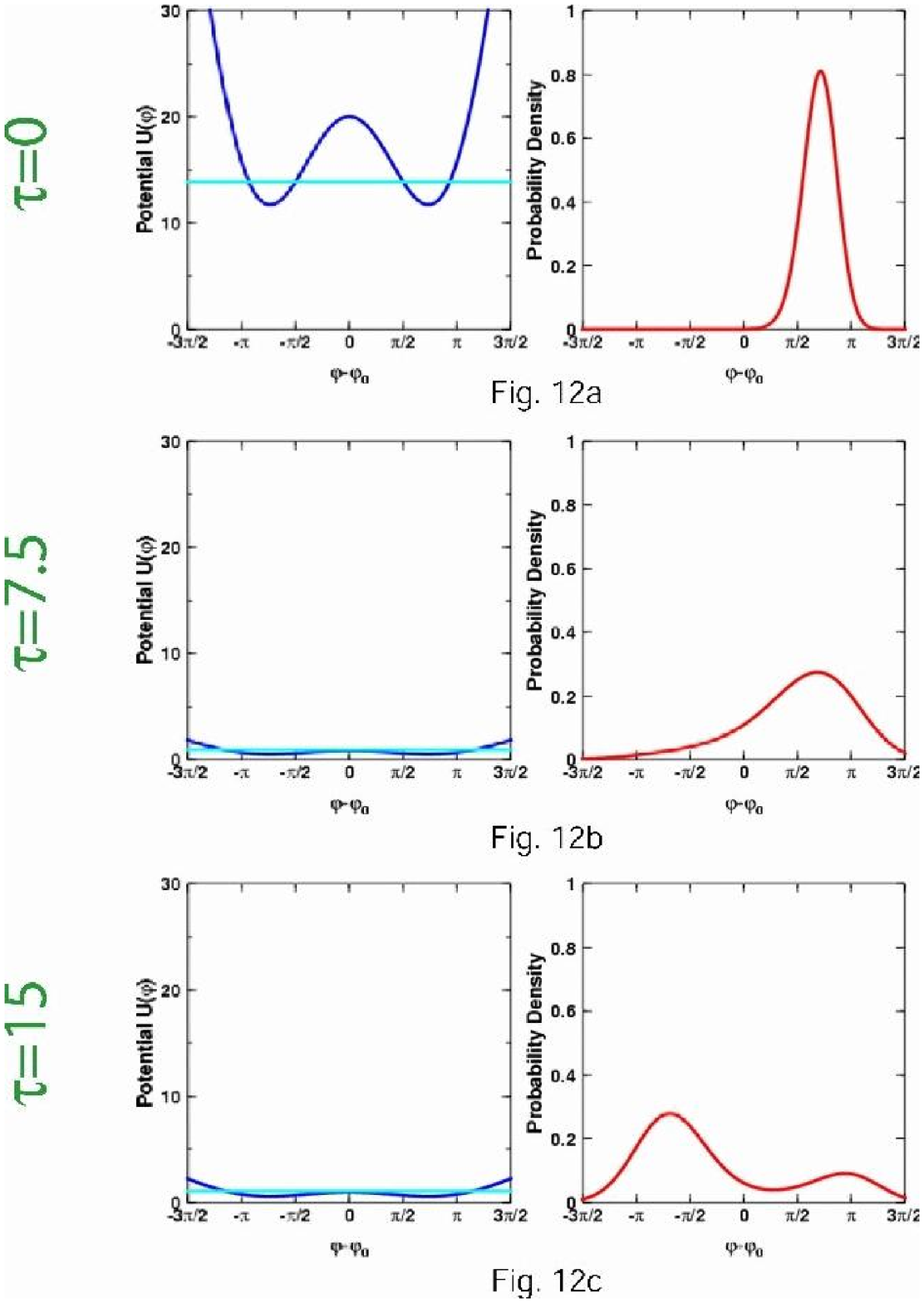}}
\endinsert

\topinsert
\centerline{
\epsfysize=7truecm \epsfbox[72 285 538 505 ]{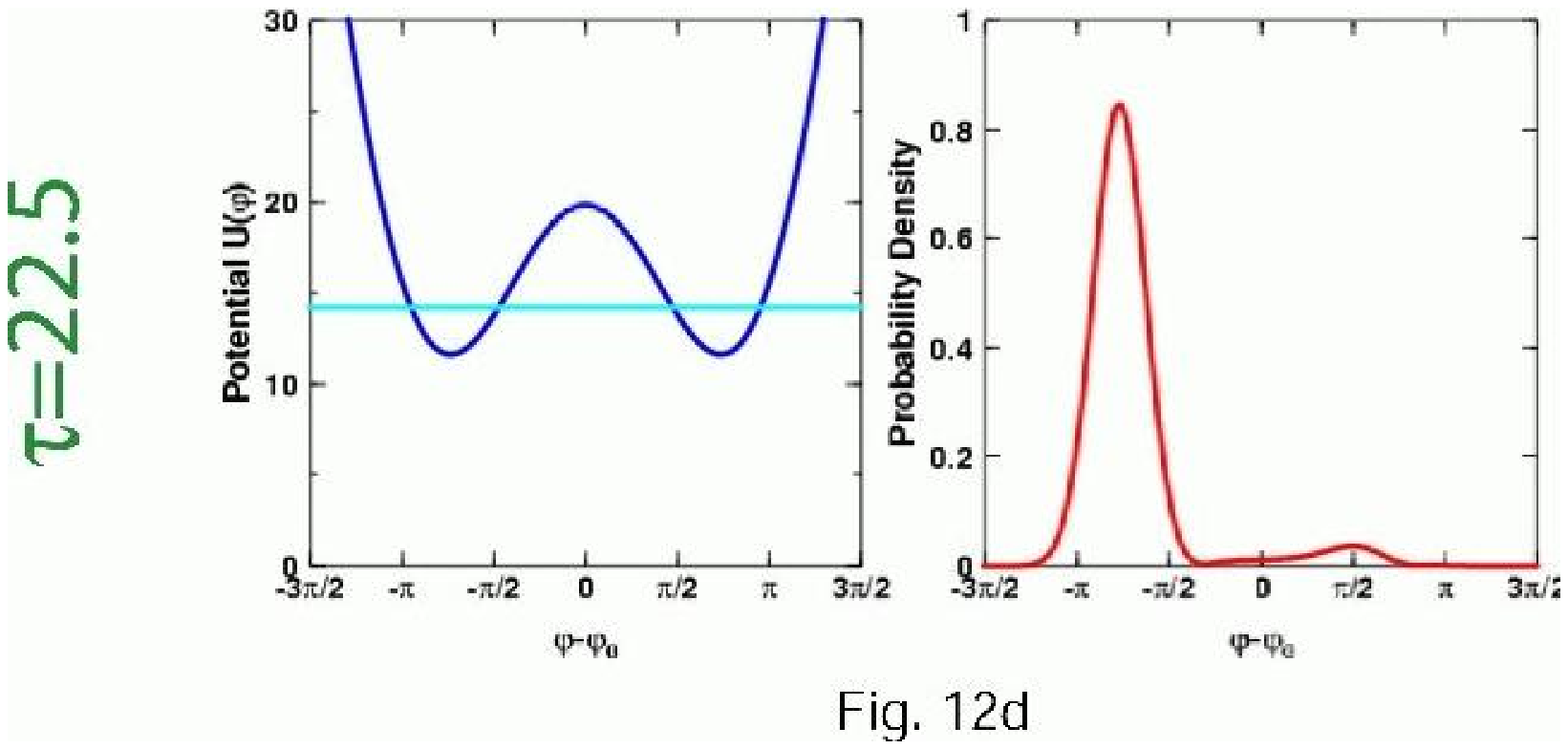}}
\bigskip
\hskip -0.6truecm\vbox{\hsize=16truecm  \noindent
Fig.~12: Snapshots of the dimensionless potential for a SQUID with
$\betaL=\phizero=\pi$ and variable $\alpha$. The light blue horizontal
lines represent the system's time dependent energy.  The evolution of the
SQUID's probability density illustrates the system's response to the
background.}
\endinsert

The energy of the initial state as well as its probability density are
graphed in fig.~12a.  As can be seen in the figure, the horizontal light
blue line representing the state's energy lies significantly below the
local center maximum of the SQUID potential.  Probability density
oscillation between the wells is classically forbidden, and it only slowly
proceeds quantum mechanically.  In order to increase the frequency of
wavefunction beating, we need to lower the potential barrier.  This
reduction may be implemented by diminishing the coupling constant.

Modulating a SQUID's $\alpha=\EJ/\EC$ value can be experimentally achieved
by subjecting the Josephson junction to an external magnetic field.  The
rate at which $\alpha$ is modified is constrained by the information theory
requirement that high energy SQUID states not be unduly excited.  In
particular, we do not want the potential's time dependence to introduce
excitations which are much more energetic than the separation
$$ \Delta E_{10} \equiv {E_1^{(+)}+E_1^{(-)} \over 2} - 
{E_0^{(+)}+E_0^{(-)} \over 2} $$
between the zeroth and first pairs of SQUID eigenstates.  This second
relevant energy scale provides an order-of-magnitude estimate $\tau_{\rm
trans} \approx 2 \pi /\Delta E_{10}=1.6$ for the time during which $\alpha$
should transition from its large initial value to a smaller intermediate
size.

We next use our simulator to determine an intermediate coupling for which
well-hopping becomes classically allowed.  We find that if $\alpha$ is
reduced from 10 to 0.4 according to the schedule shown in
\alphavariation, the resulting SQUID energy lies slightly above the
modified potential's barrier.  The dashed vertical red lines in the figure
depict the transition time $\tau_{\rm trans}$.  The potential, system
energy and system probability density after the transition is complete are
illustrated in fig.~12b.  As we see from the eigenenergies for the
$\alpha=0.4$ potential
\eqn\intermediateSQUIDspectrum{\eqalign{
\varepsilon_0^{(+)} = 0.7377   \qquad    \varepsilon_0^{(-)} = 0.9834 \cr
\varepsilon_1^{(+)} = 1.5514   \qquad    \varepsilon_1^{(-)} = 2.1090, \cr}}
the splitting between the ground and first excited state is two orders of
magnitude larger than that for its $\alpha=10$ counterpart.  So the SQUID's
probability density migrates nearly 100 times faster from one well to the
other following the transition.

\midinsert
\centerline{
\epsfysize=8truecm \epsfbox[0 0 565 440 ]{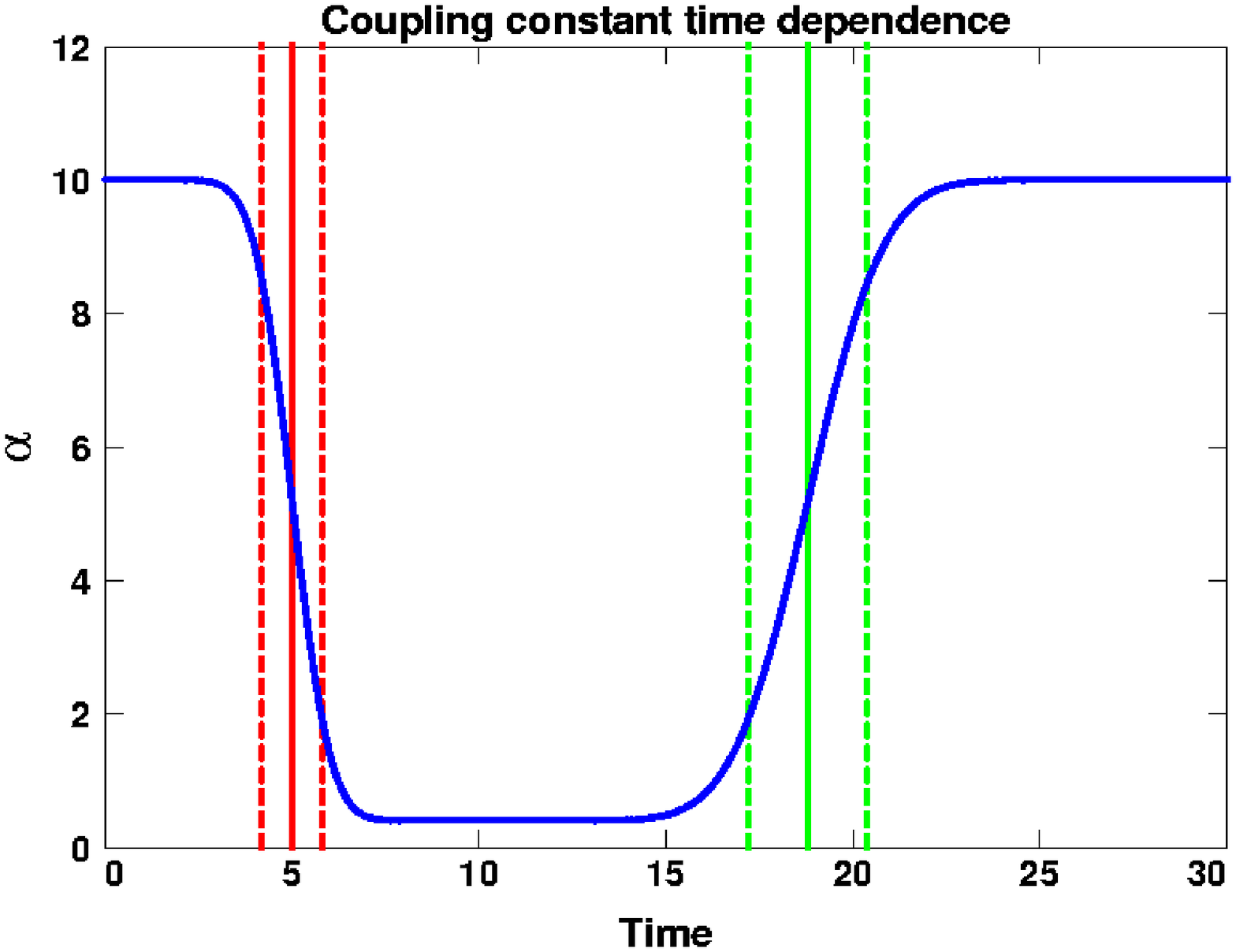}}
\bigskip
\hskip -0.6truecm\vbox{\hsize=16truecm  \noindent
Fig.~13: SQUID coupling time dependence which induces NOT gate behavior:
$$ \alpha(\tau) = 10 + {10 - 0.4 \over 2} \Bigl[
{\rm erf} \Bigl( {\tau-5 \over 0.8 \sqrt{2}} \Bigr)
- { \rm erf} \Bigl( {\tau - 13.8 \over 1.6 \sqrt{2}} \Bigr) \Bigr]. $$}
\endinsert

After an interval approximately equal to one half of the new beat period
$0.5 \, T'_{\rm beat} = \pi /(\varepsilon_0^{(-)} -
\varepsilon_0^{(+)}) = 12.8$, most of the SQUID's probability density
resides within the negative well (see fig.~12c).  We then want to raise the
barrier back to its original height in order to trap the bulk of the
wavefunction in its new location.  The separation
$$ \Delta \varepsilon_{10} ={\varepsilon_1^{(+)} + \varepsilon_1^{(-)}
\over 2} - {\varepsilon_1^{(+)} + \varepsilon_1^{(-)} \over 2} $$
between the zeroth and first pairs of $\alpha=0.4$ SQUID eigenstates again
provides an order-of-magnitude upper bound for a reasonable second
transition time $\tau'_{\rm trans} \approx 2 \pi /\Delta \varepsilon_{10} =
6.5$.  After running several simulation tests, we find that taking
$\tau'_{\rm trans}=3.2$ leads to successful wavefunction capture with a
final SQUID energy $E_{\rm final} = 14.3$ that is quite close to its
original value $E_{\rm init} = 13.9$.  The timing and duration of the
potential's return to its original form are depicted by the green lines in
\alphavariation.  The system's probability density after the entire
coherent operation is over is shown in fig.~12d.

A measurement made of the SQUID's final current sense is highly likely to
be opposite to that of its initial direction.  So driving the
$\alpha=\EJ/\EC$ coupling according to a schedule like that in
\alphavariation\ represents one possible way to implement a
quantum NOT gate.  It is interesting to note that all quantum computing
logic can be built up from NOT gates as well as CNOT (controlled not)
gates.  SQUID implementations of CNOT have also been considered in the
recent literature
\Corato.

\newsec{Conclusion}

In this article, we have developed a new approach for numerically solving
Schr\"odinger's equation, and we have used it to analyze the low energy
behavior of several quantum systems.  Our numerical algorithms are based
upon a Baker-Campbell-Hausdorff expansion of the time evolution operator
which manifestly preserves unitarity and works in any number of spatial
dimensions.  We have also identified a ratio of characteristic potential to
kinetic energies as a key coupling constant $\alpha$ that fixes the
strength of the interaction between a quantum system and its classical
background.  For problems where $\alpha$ represents the only dimensionless
parameter whose value significantly exceeds unity, dimensional analysis
establishes relevant length and time scales for low energy states.  It is
important to take the $\alpha$ dependence of these physical scales into
account when numerically integrating Schr\"odinger's equation.

We have applied our numerical techniques to compute energy eigenvalues and
eigenstates for a number of examples.  The methodology has been validated
by reproducing known answers for analytically soluble problems and
computing consistent results for analytically intractable models.  We have
also simulated the evolution of bound states and propagating packets.
Various quantum phenomena such as potential barrier tunneling, stationary
state interference, and wavefunction diffusion have been graphically
displayed.  These illustrative examples help develop quantum intuition,
especially when they clash with classical expectations.

Finally, we have demonstrated that our approach can be usefully applied to
systems interacting with time dependent potentials.  In particular, it
provides a valuable tool for designing quantum information devices.
Theoretical simulation of quantum circuits can help guide ongoing and
future experiments which are important from both fundamental physics and
technological standpoints.  Indeed, this last point provided the impetus
for this work.

In closing, we briefly mention a few applications of the methods and
insights presented in this article which we believe will be interesting to
pursue.  One current outstanding challenge in quantum computing is to build
a robust measuring apparatus that can read out a qubit's state without
prematurely destroying its quantum information.  SQUID technology is being
pursued for not only qubit construction but also measurement operations.
Preliminary simulations of such SQUID measuring circuits evince many of the
same quantum issues as the more elementary examples discussed in this
paper: $\alpha=\EJ/\EC$ values which are orders of magnitude larger than
unity, initial wavefunctions which are concentrated in deep potential
wells, temporal pulse shaping which is needed to avoid undue excitations,
time scales for wavepacket mean motion which are longer than those for
packet spreading, etc. Theoretical simulation and experimental
implementation of SQUID measurement circuits will be reported elsewhere
\Berggren.  But clearly, many of the ideas discussed here can be usefully
applied to exciting problems that lie at the quantum information processing
frontier.

\bigskip\bigskip
\centerline{\bf Acknowledgments}
\medskip
We thank A.~Bradley, T.~Imbo, M.~Misiak and W.~Oliver for their helpful
comments on the manuscript.

\vfill\eject
\appendix{A}{Position and momentum space conventions}

We collect together in this appendix various useful position and momentum
eigenstate and operator identities in $n$ spatial dimensions.  We follow
Feynman's approach to keeping track of where factors of $2 \pi$ appear in
these relations.  One factor of $2 \pi$ accompanies every momentum space
delta function in each spatial dimension, while $1 /2\pi$ accompanies every
momentum space measure factor.  No other sources of $2 \pi$ enter into the
identities below.

\bigskip\noindent
$\underline{\rm Eigenstate \; orthonormality \; relations}$:
$$ \eqalignno{
\dotproduct{\xvec}{\xpvec} &= \delta^{(n)} (\xvec - \xpvec) \cr
\dotproduct{\pvec}{\ppvec} &= (2\pi)^n \delta^{(n)} (\pvec - \ppvec) \cr} $$

\bigskip\noindent
$\underline{\rm Eigenstate \; completeness \; relations}$:
$$ \int d^nx \, \ket{\xvec} \bra{\xvec}
   = \int {d^n p \over (2 \pi)^n} \, \ket{\pvec} \bra{\pvec} 
   = {\bf 1} $$ 
\bigskip\noindent
$\underline{\rm Delta \; function \; identities}$:
$$ \eqalignno{
\int d^nx \, e^{i \xvec \cdot (\pvec - \ppvec)} &= (2\pi)^n \delta^{(n)}
(\pvec - \ppvec) \cr
\int {d^np \over (2\pi)^n} \, e^{i \pvec \cdot (\xvec - \xpvec)} &=
\delta^{(n)}(\xvec - \xpvec) \cr} $$

\bigskip\noindent
$\underline{\rm Position  \; and \; momentum \; eigenstate\; overlap}$:
$$ \dotproduct{\xvec}{\pvec} = e^{i \xvec \cdot \pvec} \qquad\qquad
\dotproduct{\pvec}{\xvec} = e^{-i \pvec \cdot \xvec} $$

\bigskip\noindent
$\underline{\rm Fourier \; transform \; conventions}$:
$$ \eqalignno{
\tilde{\psi}(\pvec) &= \int d^n x e^{-i \pvec \cdot \xvec} \psi(\xvec)  
\equiv \CF \bigl(\psi(\xvec) \bigr) \cr
\psi(\xvec) &= \int {d^np \over (2\pi)^n}  e^{i \xvec \cdot \pvec} 
\tilde{\psi}(\pvec) \equiv \CF^{-1} \bigl( \tilde{\psi}(\pvec) \bigr). \cr} $$

Here $\psi(\xvec) = \dotproduct{\xvec}{\psi}$ denotes a position space
wavefunction, while $\tilde{\psi}(\pvec) = \dotproduct{\pvec}{\psi}$
represents its momentum space counterpart.

\bigskip\noindent
$\underline{\rm Commutation \; relations}$:
\foot{Recall that the commutator of two hermitian operators always equals
an anti-hermitian combination of operators.}
$$ \eqalignno{
\bigl[ \Pvec,\Xvec \bigr] &= -i \cr
\bigl[ \Pvec,V(\Xvec) \bigr] &= -i \nabla V(\Xvec) \cr
\bigl[ \Pvec^2,V(\Xvec) \bigr] &= 2 \bigl\{ -i \nabla V(\Xvec) \bigr\}
\cdot \Pvec + (-i \nabla)^2 V(\Xvec) \cr} $$

\appendix{B}{Explicit but approximate $\pmb{O(} t^4 \pmb{)}$ 
wavefunction evolution formula}

After expanding all commutators in the time evolution operator $\bfU(t)$
shown in \BCHdecomp, we let it act upon an initial state vector
$\ket{\psi(0)}$:
$$ \eqalign{
\ket{\psi(t)} &= e^{i t^2 \vec{\nabla} \cdot \vec{\bfP}} 
e^{\bigl[ it^3 \bigl( \twothirds \vec{\nabla}V \cdot \vec{\nabla}V + \sixth
\nabla^2 \nabla^2 V \bigr) + \half t^2 \nabla^2 V - i t V \bigr]} \cr
& \quad \times e^{\bigl[ -\twothirds i t^3 \sum_j \bfP_j \nabla^2 V \bfP_j
- \fourthirds i t^3 \sum_{j > k} (\partial_j \partial_j V) \bfP_j \bfP_k
- it {\vec\bfP} \,^2 \bigr] } \> \ket{\psi(0)}.} $$
For computational speed purposes, we want to derive a wavefunction
evolution formula that involves just one Fourier transform and one inverse
transform.  So we approximate $ \sum_j \bfP_j \nabla^2 V \bfP_j \approx C
\, \vec{\bfP} \,^2 $ and 
$\sum_{j > k} (\partial_k \partial_j V) \bfP_j \bfP_k \approx
\sum_{j>k} C_{jk} \bfP_j \bfP_k$ for some potential-dependent constants $C$ 
and $C_{jk}$. We then project the state vector onto the position basis to
deduce
$$ \eqalign{
\psi(\xvec,t) & \simeq
e^{\bigl[ it^3 \bigl(\twothirds \vec\nabla V(\xvec') \cdot
\vec\nabla V(\xvec') + \sixth \nabla^2 \nabla^2 V(\xvec') \bigr) 
+ \half t^2 \nabla^2 V (\xvec') - it V (\xvec') \bigr]} \cr & \quad \times
\CF^{-1} \Bigl[ e^{-\fourthirds i t^3 \sum_{j>k} C_{jk} p_j p_k - \bigl(
\twothirds C it^3 + it \bigr) {\vec p} \,^2 }
\CF \bigl( \psi(\xvec,0) \bigr) \Bigr] + O(t^4)} $$
where ${\xvec}' = \xvec + t^2 \vec\nabla V(\xvec)$.  

\listrefs
\bye